\documentclass[aip, apl, showpacs, reprint, preprintnumbers,amsmath,amssymb,superscriptaddress,floatfix]{revtex4-1}

\usepackage[]{graphicx}      
\usepackage{bm}             	
\usepackage{times}			
\usepackage{tabularx}
\usepackage{color}
\usepackage{dcolumn}		
\usepackage{amsmath}
\usepackage{amssymb}
\usepackage{amsfonts}
\usepackage{times}
\usepackage{tabularx}
\usepackage{mathrsfs}
\usepackage{xcolor,colortbl}

\usepackage{amsmath}

\makeatletter
\newcommand\xleftrightarrow[2][]{%
  \ext@arrow 9999{\longleftrightarrowfill@}{#1}{#2}}
\newcommand\longleftrightarrowfill@{%
  \arrowfill@\leftarrow\relbar\rightarrow}
\makeatother

\begin{document}
\title{\textcolor{black}{Measuring the dispersion relations of spin wave bands using time-of-flight spectroscopy}}
\author{T. Devolder}
\email{thibaut.devolder@c2n.upsaclay.fr}
\affiliation{Universit\'e Paris-Saclay, CNRS, Centre de Nanosciences et de Nanotechnologies, 91120, Palaiseau, France}
\author{G. Talmelli}
\affiliation{imec, Kapeldreef 75, 3001 Heverlee, Belgium}
\affiliation{KU Leuven, Department of Electrical Engineering, Kasteelpark Arenberg 10, 3001 Leuven, Belgium}
\author{S. M. Ngom}
\affiliation{Universit\'e Paris-Saclay, CNRS, Centre de Nanosciences et de Nanotechnologies, 91120, Palaiseau, France}
\author{F. Ciubotaru}
\author{C. Adelmann} 
\affiliation{imec, Kapeldreef 75, 3001 Heverlee, Belgium}
\author{C. Chappert}
\affiliation{Universit\'e Paris-Saclay, CNRS, Centre de Nanosciences et de Nanotechnologies, 91120, Palaiseau, France}

\date{\today}                                           
                       
%
%
\begin{abstract}
We develop a generic all-inductive procedure to measure the band structure of spin waves in a magnetic thin stripe. In contrast to existing techniques, our method works even if several spin wave branches coexist in the investigated frequency interval, provided that the branches possess sufficiently different group velocities. We first measure the microwave scattering matrix of a network composed of distant antennas inductively coupled to the spin wave bath of the magnetic film. After a mathematical transformation to the time-domain to get the transmission impulse response, the different spin wave branches are viewed as wavepackets that reach successively the receiving antenna after different travel times. In analogy with time-of flight spectroscopy, the wavepackets are then separated by time-gating. The time-gated responses are used to recalculate the contribution of each spin wave branch to the frequency domain scattering matrix. The dispersion relation of each branch stems from the absolute phase of the time-gated transmission parameter. The spin wave wavevector can be determined unambiguously if the results for several propagation distances are combined, so as to get the dispersion relations. 
\end{abstract}

\maketitle

%
%

\section{Introduction}
The spin waves possess features --frequency tunability, short wavelengths, strong non-linearity and non-reciprocity-- that make them intriguing quasi-particles that are uniquely suited for the implementation of innovative microwave functions. 
A prerequisite for most applications that rely on spin waves \cite{Kruglyak_magnonics_2010, chumak_magnon_2015} is to know their band structure, i.e. the dispersion relation of each spin wave branch present at the frequencies of interest. 
Many of the past experimental determinations of the spin wave band structure were based on Brillouin Light Scattering (BLS) experiments \cite{hillebrands_progress_1999}, or its space-resolved variant \cite{demokritov_micro-brillouin_2008, sebastian_micro-focused_2015}. More recently, inductive microwave measurements based on Vector Network Analyzers (VNAs) have become popular to study spin waves \cite{bailleul_propagating_2003}. Coupled with the high sensitivity of VNAs, the fabrication of nanoantennas allows to reach large wavevectors\cite{ciubotaru_all_2016, talmelli_electrical_2021} , thereby providing capabilities that compare well with BLS systems. The measurement of the full spin wave band structure using all-electric means is thus now within reach. 

Attempts toward this goal are numerous in the recent literature \cite{yu_high_2012, ciubotaru_all_2016, maendl_spin_2017, qin_propagating_2018, Stuckler_ultrabroadband_2017, chen_spin_2018, qin_propagating_2018, Stuckler_ultrabroadband_2017, chen_spin_2019, sheng_spin_2020, sushruth_electrical_2020}. These experiments are all based on the measurement of spin wave propagation between antennas. The frequency dependence of the transmission coefficient can be analyzed \cite{yu_high_2012, maendl_spin_2017, qin_propagating_2018, Stuckler_ultrabroadband_2017, sheng_spin_2020} to provide the spin wave group velocity $\frac{\partial \omega} {\partial k_x}$ for a set of discrete frequencies $\frac \omega{2 \pi}$. Here $k_x$ is the wave vector along the spin wave propagation direction. Getting the true dispersion curve $\omega(k_x)$ requires to assume that the spin wave is not strongly dispersive and then to perform an integration. The integration constant is a wave vector that has to be determined. This difficulty is sometimes circumvented by postulating that the wavevectors can be exactly deduced from the antenna geometry and the profile of its rf fields \cite{chen_spin_2018, qin_propagating_2018, Stuckler_ultrabroadband_2017, chen_spin_2019, sushruth_electrical_2020}. This assumption is risky since  the antenna radiation pattern can substantially differ from analytical estimates when permeable materials with finite conductivity are present in the surroundings of the antenna \cite{bailleul_shielding_2013}. Equivalently, the frequency dependence of the transmission parameter is sometimes fitted directly to the theoretical dispersion curve \cite{qin_propagating_2018, talmelli_reconfigurable_2020} using a single-mode signal theory \cite{sushruth_electrical_2020}.

These previous methods suffer from two important limitations. The first one is the inability to directly measure the dispersion relation, or to deduce it without making uncheckable assumptions. The second limitation is that these methods require that a \textit{single} spin wave branch contributes to most of the signal amplitude in the investigated frequency interval: these methods are bound to fail if the spin wave density of states comprises several branches with substantial contributions. Our present paper aims at developing an enhanced method that solves these two issues in order to construct the spin wave band structure in an indisputable manner. Of course this can be done --and has been done routinely-- in the past by the BLS community \cite{demidov_magnonic_2015}; our present purpose is to study how to do it in an all-electrical manner so as to benefit from the very broad frequency coverage of VNAs as well as their exceptional dynamic ranges that permit fast measurements. Since our goal is to discuss a novel methodology, we shall implement it on a model system whose spin wave spectrum is well known\cite{bayer_spin-wave_2005, demidov_magnonic_2015, talmelli_electrical_2021}: a micrometer-sized stripe magnetized along its width. 

The paper is organized as follows. We first describe the sample design in section \ref{design}. The sample is chosen to reveal both the potential of our technique and its main hurdles. The detailed experimental methodology and the results recorded in frequency domains are reported in section \ref{exp}. Section \ref{interconversion} then describes how to mathematically transform the frequency domain data into time-domain impulse responses to perform time-of-flight spin-wave spectroscopy. When there are several spin wave branches, time-gating is performed to separate their contributions to the total density of states. The dispersion relations are then constructed in section \ref{constructionOfDispersion}. As a final check, these dispersions are compared to expected ones, and are used to identify the experimental efficiency spectrum of the antenna.

\section{Sample design for time-of-flight spin-wave spectroscopy} \label{design}
\subsection{Objectives and design strategy} \label{design}
\begin{figure*}
\hspace*{-0.1cm}\includegraphics[width=16 cm]{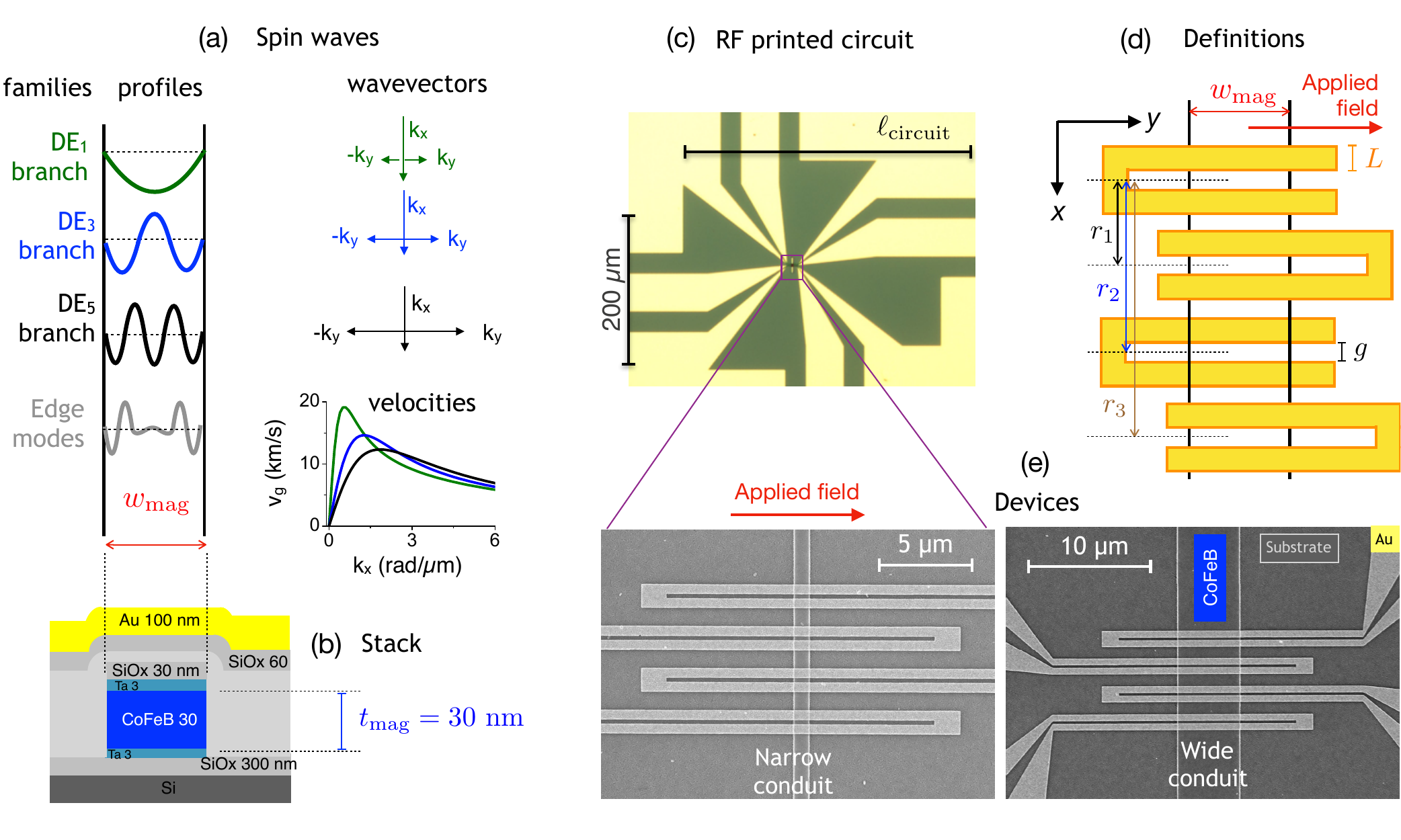}{\centering}
\caption{(a) Sketch of the different families of spin wave branches, their spatial profiles within the conduit width, their effective wavevectors and their group velocities for $w_\textrm{mag}=4.7~\mu\textrm{m}$ at 18 mT. (b) Cross section of the sample, with thicknesses in nm. (c) Far field optical view of the rf contacts routing the electrical signals towards the antenna. The metallic parts appear in golden color. (d) Sketch of the device and definitions of its dimensions. (e): Scanning electron micrographs of the central parts of devices with conduit widths $w_\textrm{mag}=4.7~\mu\textrm{m}$ (left) and $w_\textrm{mag}= 0.85~\mu\textrm{m}$ (right).}
\label{fig1}
\end{figure*}

Our method will rely on the detailed analysis of the phase of spin waves after their propagation through a magnetic medium. Inductive antennas are used as spin wave emitters and receivers [Fig.~\ref{fig1}], in line with what is commonly practiced for all-electrical spin wave analysis \cite{vlaminck_current-induced_2008, vlaminck_spin-wave_2010, gladii_spin_2016, ciubotaru_all_2016, talmelli_reconfigurable_2020, sushruth_electrical_2020, bhaskar_backward_2020}. Since we aim to determine the dispersion relation, we use narrow antennas to cover an as-large-as-possible interval of wavevectors\cite{sushruth_electrical_2020}. We have implemented our time-of-flight spin-wave spectroscopy method on several types of samples and of spin waves ; for a didactic purpose, we illustrate it here with samples that best reveal the potential of this technique. This requires that many families of spin wave modes coexist at the same frequency and propagate in the same direction.

This condition is conveniently obtained in the Damon-Eshbach configuration: we harness the so-called Magneto-static Surface Spin Waves (MSSW) that propagate along the length $x$ of a narrow magnetic conduit of finite width $w_\textrm{mag}$ and thickness $t_\textrm{mag}$ much smaller than $w_\textrm{mag}$, submitted to a transverse field $H_y$ that saturates the magnetization [Fig.~\ref{fig1}(a)]. The lateral confinement within the magnetic waveguide forces the spin waves \cite{bayer_spin-wave_2005} to get a standing wave character in the transverse direction ($y$) while the longitudinal wavevector $k_x$ remains free to take any value. As shown by V. Demidov et al. \cite{demidov_mode_2008, demidov_magnonic_2015}, this naturally creates several families of spin waves with different dispersion relations in the $k_x$ (propagative) direction.

A first family of spin wave branches is the confined version of the plane waves existing in the extended films. Following the convention of ref. \cite{demidov_magnonic_2015}, we index them as DE$_1$, DE$_2$ ,..., where the superscript $m=1,2...$ is the number of antinodes in the width of the magnetic conduit, which can be viewed as an effective quantization of the transverse wavevector at values of $k_y=\frac{m \pi}{w_\textrm{mag}}$ [see the sketches in Fig.~\ref{fig1}(a)]. Among these branches of modes, the ones with odd $m$ are excitable and detectable by inductive antennas\cite{bracher_creation_2017}, with an efficiency essentially scaling like $1/m$; they will be our primary Guinea pigs in this paper. \textcolor{black}{A convenient feature of these confined modes is that as the index $m$ increases, the modes progressively loose their MSSW character (i.e. with a large group velocity at low $k_x$ and an effective wavevector $\vec k_x +\vec  k_y$ almost along the equilibrium magnetization) to resemble more and more the magneto-static backward volume spin waves (MSBVSW) with much smaller group velocities \cite{slavin_nonlinear_2002} and effective wavevectors almost perpendicular to the equilibrium magnetization}. In the saturated state, \textcolor{black}{the exchange contributions can be neglected at small wavevectors such that the spin wave} dispersion relation can be approximated by \cite{demidov_magnonic_2015}: 
\begin{equation} \omega_\textrm{DEm}(k_x)=\gamma_0 \sqrt{H_1 H_2} \label{DE} \end{equation} where the stiffness fields are $H_1=H_0 + M_s (1-P)$ and $H_2=H_0 + M_s (P \frac{k_x^2}{k^2})$, with $P=1+\frac{e^{-k t_\textrm{mag}}-1}{k t_\textrm{mag}}$,  \textcolor{black}{ $k^2=k_x^2+k_y^2$} 
and $H_0$ being the internal magnetic field \textcolor{black}{(i.e. including the applied field as well as the demagnetizing field related to the shape anisotropy)} and $M_s$ the magnetization. (Note that in ref.~\cite{demidov_magnonic_2015} the applied field $H_y$ and the internal field $H_0$ were taken as equal because the demagnetizing effects were small; this is not the case here because we use materials with much larger saturation magnetization.) 
This expression will be used to confirm the experimentally determined dispersion relations.

The dispersion relation ressembles that of the extended film only when $k_x w_\textrm{mag} \gg m \pi $. At $k_x=0$, the lateral confinement distorts the dispersion relation which gets a vanishing group velocity  [Fig.~\ref{fig1}(a)]. As a result, the spin wave attenuation lengths $L_\textrm{att}$ also vanish at $k_x=0$: these spin waves cannot reach the receiver antenna. It is important to figure out that when propagating spin wave spectroscopy is conducted in narrow conduit, the vicinity of the $k_x=0$ points of the DE$_m$ branches are intrinsically not measurable.
%

In addition to this DE$_m$ family of spin waves, edge modes can exist at fields slightly above the one needed to saturate the magnetization of the stripe along its width \cite{bayer_spin-wave_2005}. 
These modes [Fig.~\ref{fig1}(a)] have no analog in the infinite film limit, and their frequencies cannot be accounted for analytically. 
Since these modes occupy only a small fraction of the stripe volume, they couple weakly to inductive antennas; besides, their low group velocities lead to a strong attenuation upon propagation;  we will see that we can anayway still measure their dispersion relation.  

\subsection{Sample geometry and properties of the magnetic material}
Let us now describe our samples and define our notations. We use Ta/CoFeB /Ta films [Fig.~1(a)] of thickness $t_\textrm{mag}$=30 nm, magnetization $M_s=1340~\textrm{kA/m}$ and Gilbert damping $\alpha\approx 0.004$. They have essentially isotropic properties. We will typically work with applied fields that lead to ferromagnetic resonance frequencies $\omega_\textrm{FMR}/ 2\pi$ in the range of 5 GHz with FMR linewidth $\Delta f \approx 200~\textrm{MHz}$. For this FMR frequency, the group velocities in the unpatterned film at $\vec k \approx \vec 0$ would be \cite{yu_high_2012}: $\frac{\partial \omega}{\partial k_x}=    \gamma_0^2 M_s^2 t_\textrm{mag} / (4 \omega_\textrm{FMR}) =\textrm{20.9~km/s}$ in the MagnetoStatic Surface Wave configuration when $\vec{k} \perp \vec{M}$, and $\frac{\partial \omega}{\partial k_y}= - \gamma_0^2 H_y M_s t_\textrm{mag} / (4 \omega_\textrm{FMR})=-216~\textrm{m/s}$  in the Backward Volume Spin Wave configuration when $\vec{k} \parallel \vec{M}$. 

The CoFeB films are patterned into magnetic conduits of widths $w_\textrm{mag} = 4.7$ and $0.85 ~\mu \textrm{m}$ and much longer lengths [Fig.~\ref{fig1}(d)]. The length and the width define the $(x)$ and $(y)$ directions. The conduits are submitted to transverse fields $H_y$ that exceed the dipolar shape anisotropy fields of $\mu_0 H_\textrm{sat}$ of the two conduits, respectively 6 mT (wide conduit) and 30 mT (narrow conduit). The applied fields are thus meant to saturate the magnetization along the width direction.
In order to reach high wavevectors, we use U-shaped antenna with narrow gaps of $g=231~\textrm{nm}$ and arm widths of $L=513~\textrm{nm}$. The antenna are placed above the CoFeB spin wave conduits at antenna-center to antenna-center being $r_1=2.3~\mu\textrm{m}$, $r_2=4.6~\mu\textrm{m}$ and $r_3=6.9~\mu\textrm{m}$ [see Fig.~\ref{fig1}(d)]. 
\section{Frequency domain experiments} \label{exp}
The spin wave time-of-flight spectroscopy is based on frequency-resolved spin-wave transmission characterizations, as detailed below.

\subsection{Specific methodology for the acquisition of the frequency domain data}
The devices are characterized by measuring their scattering matrix $\tilde{S}_{ij}$ with $i, j  \in \{1, 2\}$ (the tilde recalls that it is a complex-valued function) with a vector network analyser (VNA) and an \textit{rf} probe station, versus frequency and versus external applied field. \textcolor{black}{The VNA output power was set to 0 dBm; the results were checked to be independent from the chosen power when decreasing it from this value.} chosen The choice of the frequency settings (100 MHz-30 GHz with steps of 10 MHz) will be discussed later. The field is transversal to the spin-wave conduit length [Fig.~\ref{fig1}(d)]. Several experimental precautions are taken. \\
First, an on-chip full 2-port calibration with a Load-Match-Reflect standard impedance calibration kit is done to correct for the imperfections of the VNA, the cable assembly and the RF probes. This sets the zero phase (reference) planes at the device contact pads. \\
Second, the device contact pads are positioned as close as possible to the inductive antennas (printed circuitry length $\ell_\textrm{circuit} \approx 300~\mu m$) so as to limit the phase accumulated by the voltage waves along this circuitry. To avoid the need of deembedding, this accumulated phase should be negligible compared to the phase that will be accumulated by the spin waves along their propagation between the antennas. This is ensured by the condition: 
\begin{equation} 
\frac{r}{v_{g}^\textrm{sw}} \gg  \frac{\ell_\textrm{circuit}}{c_\textrm{g}^\textrm{em}} \approx 1.5~\textrm{ps}, \label{eqell}
\end{equation} 
where $r \in [2.3-6.9~\mu \textrm{m}]$ is the propagation distance of the spin waves, $v_g^\textrm{sw} \approx 2-20~\textrm{km/s}$ their group velocity and $c_\textrm{g}^\textrm{em} \approx 2\times 10^8~\textrm{m/s}$ the group velocity of the voltage waves in the rf circuitry at the frequencies of interest. We will see that the duration ${r}/{v_{g}^\textrm{sw}}$ of the travel of the spin waves between the antennas lies in the ns range, such that de-embedding of the $\ell_\textrm{circuit}$-long section is not necessary in our case. 
\\
Third, we correct for signals that are not originating from spin-wave signals. In addition to the spin-wave-mediated signals, the raw scattering matrix comprises a direct antenna reflection and direct antenna-to-antenna coupling from unavoidable inductive and capacitive effects. Several methods can be used to construct an approximation of the scattering matrix of this parasitic transmission \cite{ciubotaru_all_2016} and subtract it from the data. Here we have used the zero field $\tilde S_{ij}$ spectra: at this field, the magnetization lies along the conduit length, and in this BVSW configuration the transmission spin wave signals are vanishingly low \cite{bhaskar_backward_2020}, due to a very low excitation efficiency of inductive antennas as well as the very short spin wave attenuation length due to the very low group velocity. We thus perform the following correction:
\begin{equation} \tilde S^{\textrm{corrected}} =  \Big[ \tilde S (H_y) - \tilde S (||\vec H|| =0)  \Big] \label{corrected} \end{equation} 
We shall omit the superscript "corrected" in the remainder of this paper as all data presented from this point onward are systematically corrected. Other methods of correction could have been used, for instance by replacing the zero field spectrum by another spectrum recorded at very large fields.  The detail of this correction does not alter our forthcoming conclusions.

%
\begin{figure*}
\includegraphics[width=17 cm]{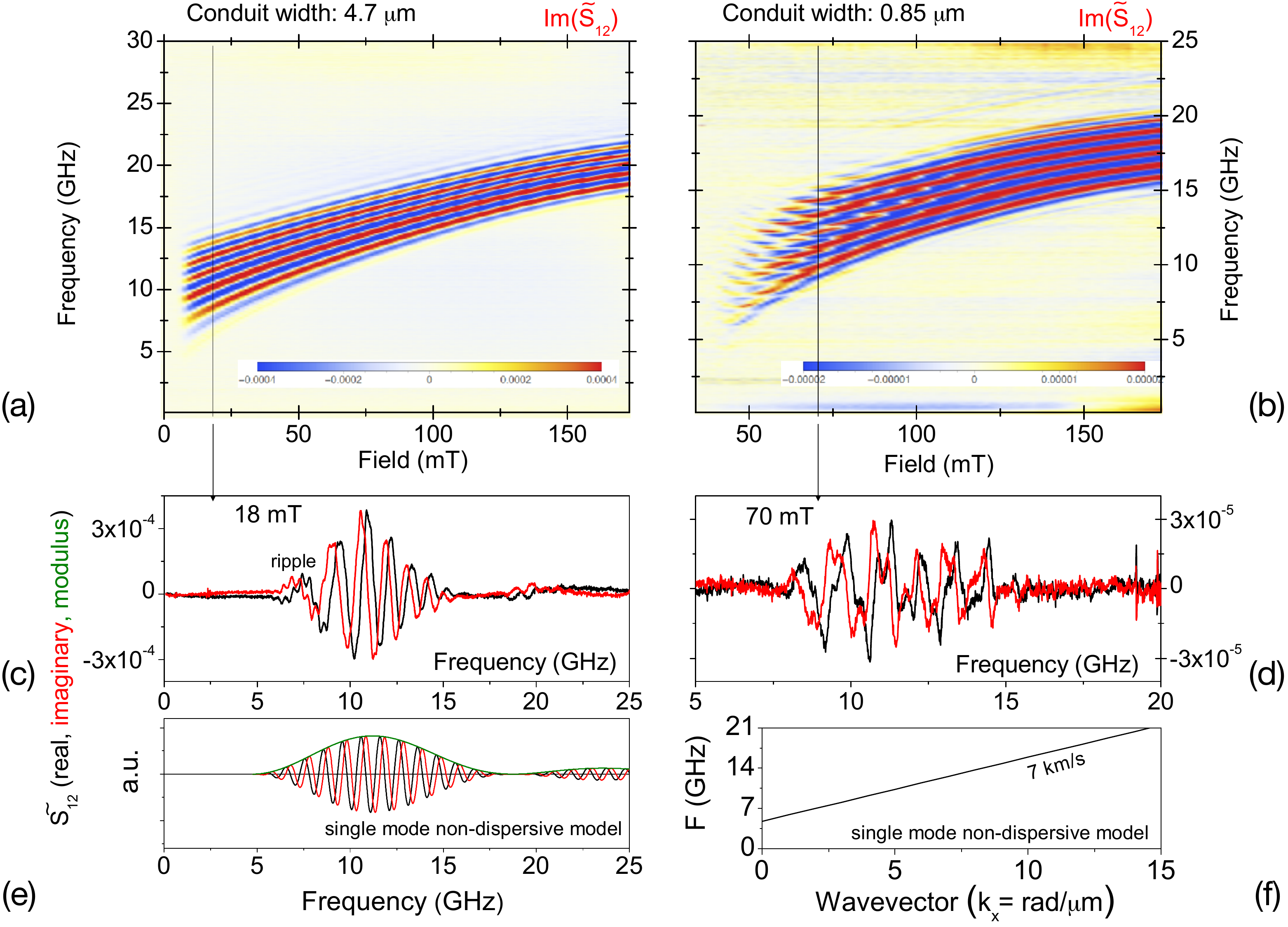}
\caption{Backward transmission parameters $\tilde{S}_{12}$ for spin-wave conduits of widths of $4.7~\mu\textrm{m}$ (left panels) and  $0.85~\mu\textrm{m}$ (right panels) for an antenna-to-antenna distance of $r_3=6.9~\mu \textrm{m}$. (a) and (b): field dependence of the imaginary part. (c) and (d): selected spectra where the multimode character is striking to the eye, either manifest as a ripple superimposed on the main enveloppe signal in (c) or as a kind of two-tone beating in (d). (e) Example of complex transmission parameter in the single mode model (Eq.~\ref{1DmonomodeTheory}), evaluated with vanishing damping for the dispersion relation displayed in panel (f). The dispersion relation starts at $f_\textrm{k=0}= 4.75~\textrm{GHz}$ and grows with a constant (hence non-dispersive) group velocity $v_g=7~\textrm{km/s}$. The envelope (green curve) in (e) is the antenna efficiency function ( Eq.~\ref{AntennaEfficiencyFunction}).} 
\label{FreqData} \label{fig2}
\end{figure*}

\subsection{Frequency domain results} 
The field and frequency dependences of the transmission parameters are reported in Fig.~\ref{FreqData} for the longest propagation distance $r=6.9~\mu \textrm{m}$. To discuss these results, it is useful to recall the transmission parameter \cite{sushruth_electrical_2020} expected when it is assumed that (i) there is a single spin-wave branch of dispersion $\omega(k_x)$ and (ii) that its attenuation length $L_\textrm{att}$ greatly exceeds the antenna dimensions $L$ and $g$ for all $k_x$ values. In this case, the signal would be a sinusoid $e^{-i k_x |r|}$ damped by the spin wave attenuation $e^{-\frac{|r|}{L_\textrm{att}}}$ and caped in an enveloppe $h_{x}^2(k_x)$ that starts from the bottom of the spin-wave band \cite{sushruth_electrical_2020} : 
\begin{equation}
    \tilde{S}_{ij}^\textrm{single~mode}(\omega) \propto  ~i ~e^{-i k_x |r|} ~e^{-\frac{|r|}{L_\textrm{att}}} ~(h_{x}(k_x))^2
    \label{1DmonomodeTheory}
\end{equation}
where the antenna efficiency function $h_{x}(k_x)$ is determined by the antenna geometry and can be semi-quantitatively described by~\cite{sushruth_electrical_2020}:
\begin{equation}
	h_{x}(k_x) \propto \textrm{sin}\!\left( \frac{k_x(g+L)}{2} \right) \frac{\textrm{sin}(k_x L/2)}{k_x L/2},
	\label{AntennaEfficiencyFunction}
\end{equation}
An example of transmission parameter expected in this single mode theory (Eq.~\ref{1DmonomodeTheory}) is given in Fig.~\ref{FreqData}(f). This idealized transmission parameter is calculated in the absence of loss (i.e. $L_\textrm{att}=\infty$) for a mode whose dispersion relation is plotted in Fig.~\ref{FreqData}(f). \textcolor{black}{This idealized and unphysical model is displayed for illustration purpose, only to ease the discussion of the later coming experimental data. } It is taken as non-dispersive, i.e. with a constant group velocity $v_g=7~\textrm{km/s}$. In this model, no spin-wave related signal is expected at frequencies below the bottom of the spin wave band; above this threshold, the phase of the transmission signals should rotate at a pace given by the dispersion relation. If the antenna is efficient for large wavevectors, a second lobe should be observed at larger frequencies, as shown in the idealized example of Fig.~\ref{FreqData}(e). 

At first glance, our results for the widest spin wave conduit [Fig.~\ref{FreqData}(a) and (c)] could seem to fall in line with the expectations of the single mode model, with the second lobe clearly visible near 20 GHz in the example of Fig.~\ref{FreqData}(c). However a closer look evidences differences. In particular, a small ripple is present at the lowest frequencies. This ripple is present only at applied fields near the saturation field of the stripe.

The difference between the single mode expectation [Fig.~\ref{FreqData}(e)] and the experimental behavior is much more striking for the narrow spin wave conduit [Fig.~\ref{FreqData}(b) and (d)], especially at low fields where the signal gives the impression of being a two-tone beating. In the single mode model the phase rotates monotonously as the frequency increases Fig.~\ref{FreqData}(e)], in contrast with the experimental result Fig.~\ref{FreqData}(d)]. Besides, the signal envelope departs substantially from the expected shape. We will see that this is the result of the presence of several spin wave branches that contribute with comparable amplitudes to the total transmitted signal. 

Thanks to their different group velocities, the separation of the different families and branches of modes is conveniently done in the time-domain by time-of-flight spectroscopy. In the next section, we use the VNA experimental data to compute how spin-wave wavepackets propagate and disperse with time in our devices.


\section{Frequency-time interconversion of spin wave signals} \label{interconversion}

The objective of this section is to calculate the device impulse response, i.e. the voltage waveform that would be measured by an oscilloscope at the receiving antenna if a voltage Dirac impulse was applied at the other antenna at the time origin $t=0$. The time $t$ is thus the \textit{travel time} of the spin wave between the emitting antenna and the receiving antenna. We use the following writing convention: frequency-domain (respectively time-domain) quantities are written in capital (resp. lowercase) letter. Complex-valued (resp. real-valued) functions are written with (resp. without) tilde. For instance, we wrote $\tilde{S}_{ij}(f)$ the frequency-dependent element of the scattering matrix of the device (i.e. the VNA data) when the power is applied at port $j \in{1, 2}$ and collected at port $i \in{1, 2}$ of the device. In a similar way, we shall write the $s_{ij}(t)$ the real-valued time-resolved voltage that would be measured by an oscilloscope at the port $i$ of the device if a voltage impulse was applied at the port $j$ at $t=0$. 

\subsection{Frequency sampling settings} 
\subsubsection{List of frequencies}
The physical signal $\tilde{S}_{ij}(f)$ is a continuous function of the frequency. However when recorded with the VNA, it is sampled at a list of $N_\textrm{points}$ discrete positive frequencies. For convenient mathematical treatment it is best to use sampled frequencies that are harmonically related, i.e. 
 \begin{equation}\{ f_\textrm{min}=\delta f,~2\delta f,~3\delta f,~...~, f_\textrm{max}=N_\textrm{points} \delta f \} \label{freqlist}\end{equation}
The "start frequency" of the VNA $f_\textrm{min}$ should thus ideally be taken equal to the frequency spacing between successive data points $\delta f$. (Note that $\delta f$ should not be confused with the ferromagnetic resonance linewidth $\Delta f$). 
\subsubsection{Negative frequency reconstruction} 
The back-and-forth translation between time and frequency domains is performed using Fourier transformation. 
The impulse responses $s_{ij}(t \geq 0)$ are real-valued (voltage) functions, so their corresponding spectra $\tilde{S}_{ij}(f)$ must be Hermitian. The negative frequency points should thus be constructed as the conjugate symmetric of the (measured) positive frequency points. The zero frequency point must also be created. As spin wave devices always rely on stable magnetic configurations, the spin wave spectrum cannot contain modes at strictly zero frequency. We can thus systematically assume that the SW contribution to the transmission and reflection parameters of the device at zero frequency vanishes. To summarize, we complete the VNA experimental dataset to get $2N_\textrm{points}+1$ frequencies by setting: \begin{equation}\tilde S_{ij} (-f) = \tilde S_{ij}^* (f)~~\textrm{and}~~ \tilde{S}_{ij} (f=0) = 0,\label{ToHermitian} \end{equation}
where the star symbol means complex conjugate.

%
\begin{figure*}
\includegraphics[width=18 cm]{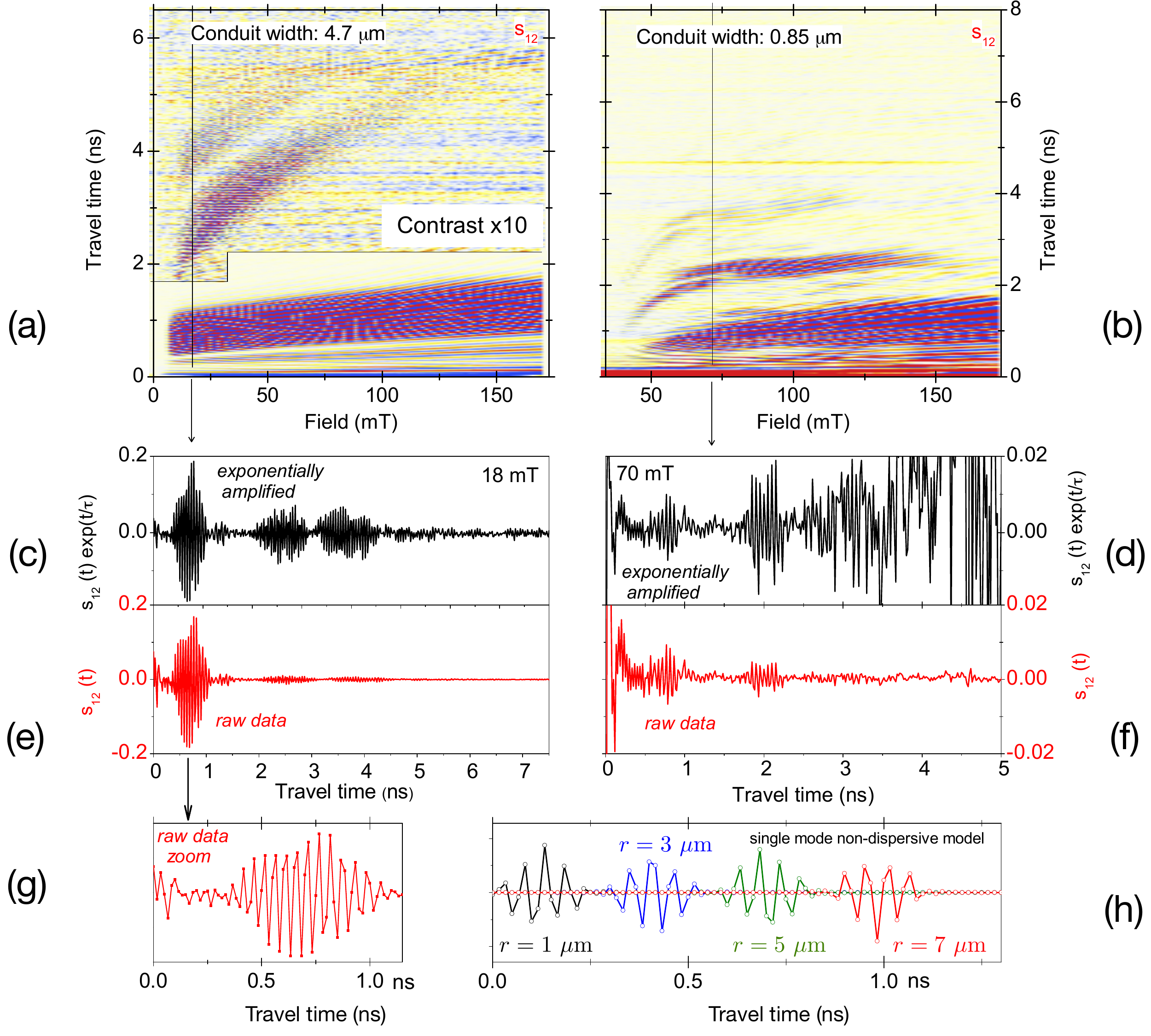}
\caption{ Backward transmission impulse response ${s}_{12}(t)$ for spin-wave conduits of widths of $4.7~\mu\textrm{m}$ (left panels) and  $0.85~\mu\textrm{m}$ (right panels) for a propagation distance $r_3=6.9~\mu \textrm{m}$. Panels (a) and (b): field dependence of the impulse response (note the change in color scale on the upper side of panel (a) that is necessary to better reveal the late-arriving wavepackets. Panels (c, d, e, f, g): selected impulse responses for which the multimode character is visible from the successive arrival of several wavepackets. The red curves are the raw data ${s}_{12}(t)$ and the black curves are time-amplified version thereof being ${s}_{12}(t) \times e^{t/\tau}$, where the amplification is meant to correct for the time-decay of the spin waves at the theoretical rate $\tau= 2 /\Delta f  \approx 1~ \textrm{ns}$. Panel (h): calculated impulse responses for several propagation distances for a theoretical loss-less spin wave mode with the dispersion relation of Fig. 2(f). }
\label{fig3}
\end{figure*}
\subsubsection{Truncation and extrapolation of the frequency spectrum} 
A word of caution is needed about frequency truncation: in some cases several of first lowest and/or last highest frequencies within the list of Eq.~\ref{freqlist} cannot be measured (for instance for instrumental limitations) or should not be measured (for instance because they only contain parasitics radiated by wireless devices). If the spin wave signal does not span over these problematic frequencies, zero padding can be used to complete the data set and recover the situation of Eq.~\ref{freqlist}.

If on the contrary the spin wave signal is not fully contained in the accessible frequency interval, the truncation of the spectrum would generate overshoots and ringing artefacts in the forthcoming time-domain data, with $(sin(t)/t)$-like features convoluting the physical signals and potentially obscuring the device response behind mathematical artefacts. Since spin wave signals in the time domain are essentially oscillatory, they could be easily confused with ringing artefacts, so truncating a part of the frequency interval in which the spin waves respond should definitely not be practiced in spin wave spectroscopy. Besides, the truncation would remove power from the physical spectrum, so that data normalization would be compromised. 

Fortunately for typical spin wave devices, the spin wave transceivers generally only emit and collect in a given spin wave wavevector interval, with steep roll-off above \cite{sushruth_electrical_2020}. Owing to the spin wave dispersion relation, this ensures a steep cut-off of the spin wave signal at high frequencies. An abrupt high frequency truncation above the usable spin-wave band thus does not impact the signal integrity. It may even reduce the noise in the time-domain impulse response, as will be shown below (see Fig.~\ref{fig4}, panels (a) and (b)).

\subsubsection{Fourier transformation to time-domain data}
Provided that the spin waves are excited in the linear regime, the impulse response can be calculated by a simple Fourier transformation of the corrected (Eq.~\ref{corrected}) and Hermitian-completed (Eq.~\ref{ToHermitian}) VNA data:
\begin{equation} s_{ij}(t \geq 0) =  \mathscr{F} \{ \tilde{S}_{ij} (f) \} \label{FFT} \end{equation}
Since the frequency data was sampled, the impulse response is also sampled and takes $2N_\textrm{points}+1$ values at harmonically-related time instants that are: 
\begin{equation}\{0, ~t_\textrm{resolution}=\frac{1}{2 f_\textrm{max}},~\frac{2}{2 f_\textrm{max}},...,~t_\textrm{max}=\frac{1}{\delta f} \}\end{equation}
The time resolution $1/ (2f_\textrm{max})$ (here: 17 ps) is the duration (FWHM) of the voltage Dirac peak that would induce the impulse response $s_{ij}(t)$.
By construction, the time domain response repeats itself every $t_\textrm{max}=1/\delta f$ after the maximum time range $t_\textrm{max}$ (here: 100 ns). This has practical consequences: if the device was such that the transmission $(i \neq j)$ impulse response would arrive after a wave travelling time $t_\textrm{travel}$ (or a two-way echo time if in reflection with $i=j$) greater than the maximum accessible time range of the experiment, then the mathematical procedure of Eq.~\ref{FFT} would alias the corresponding signal fictitiously within the accessible time range, at $t_\textrm{travel}-t_\textrm{max}$. A good way to check that this situation is not encountered is to verify that a change of $\delta f$ (hence of $t_\textrm{max}$) does not change the time-domain data.

SW signals attenuate exponentially at a time rate $\tau=2/ \Delta f$ (here $\Delta f=220$ MHz at 5 GHz for the ferromagnetic resonance), such that choosing a frequency resolution of $10~\textrm{MHz}$ is more than enough to ensure $\delta f \ll \Delta f$ and thus to guaranty an alias-free time range of the spin wave signals.


\subsection{Results of time-of-flight spin wave spectroscopy} 
The impulse responses in transmission resulting from Eq.~\ref{FFT} are displayed in Fig.~\ref{fig3} for the exact same sets of data as their frequency counterparts formerly displayed in Fig.~\ref{fig2}. This is done for the experimental data (panels (a) to (g) and the theoretical response of a single non-dispersive mode [panel (h)].

At the time origin and soon after, a large glitch is always present in the experimental impulse responses. The amplitude and the overall shape of these glitches are strongly dependent on the way the correction (Eq.~\ref{corrected} or qualitatively similar options) is implemented, which indicates that it is a residue of the imperfect subtraction of the so-called "feed-through" signal that comes from the direct coupling between the input and output antennas by capacitive and inductive effects. Since this initial part of the experimental time-domain signal is unreliable, we shall disregard it for the discussion of the spin wave properties. 

Later in time, the receiving antennas see the arrival of several successive wavepackets. The travel times of the wavepackets vary with the magnetic field [Fig.~\ref{fig3}(a-b)] and increase with the propagation distance $r$ (not shown). With our signal to noise ratio, up to three successive wavepackets can be perceived in the experiments: the fast (first arriving), the medium and the low (third arriving) velocity wavepackets. In the wide spin wave conduit case, the amplitudes of the second and the third arriving wavepackets are so low that contrast rescaling [Fig.~\ref{fig3}(a)] or mathematical compensation of the $e
^{- t \Delta f /2 }$ decay rate of the spin waves  [Fig.~\ref{fig3}(c)] is needed to evidence these wavepackets.

Compared to the later arriving wavepackets, the first arriving one has a relatively large duration, which is an indication that the group velocity of the corresponding mode has a larger frequency dependence, \textcolor{black}{i.e. it is more dispersive}. This can also be inferred from the comparison with the (idealized) wavepackets calculated for the single-mode non-dispersive model : in this case the wavepackets would not spread upon propagation [Fig.~\ref{fig3}(h)].
Because of this dispersive character of the fastest mode in the widest spin wave conduit, its group velocity cannot be deduced exactly from the time-domain data by a naive distance/time division. This naive calculation would give $6.9 ~\mu\textrm{m}/0.7 \textrm{ns} \approx 10
~\textrm{km/s}$ while we will see that it varies from 14 to 5 km/s in the investigated interval of wavevectors. In contrast, the calculation of the group velocity by distance/time division would be perfectly legitimate for a non-dispersive mode: the (hypothetical) distance of $7 ~\mu\textrm{m}$ is travelled in exactly 1 ns for the idealized non-dispersive mode with $7~\textrm{km/s}$ of group velocity [Fig.~\ref{fig3}(h)].

Coming back to the experimental data in the narrow spin wave conduit, the signal is one order of magnitude weaker but strong enough to assess that the first two wavepackets have comparable amplitudes, while the third one is much weaker. Their widths in time indicate a less dispersive character than the fastest mode of the wide conduit. In all cases the successive wavepackets arrive at the receiving antenna after very different trip durations, such that they correspond to branches of modes with different group velocities, as targeted. In the next section we will show how we can separate these branches to deduce the individual dispersion relations. Before that, we will comment on the impulse responses of the device reflection parameters $\tilde{S}_{11}$ and $\tilde{S}_{22}$. 

Indeed the impulse response can also be calculated for the reflection coefficients (not shown). In this case all spin-waves are excited at $t=0$ and their ringing immediately contribute inductively to the reflected impulse response, essentially with precessions that last from the time origin and a few $\tau=2/ \Delta f$, regardless of their group velocity. This situation is the equivalent of the pulse-induced magnetometer popularized by T. Silva et al. a while ago \cite{silva_inductive_1999} but now with the capability to excite wavevectors in a broad interval. Since the spin wave response is very broadband and spans over $\approx 10~\textrm{GHz}$ in frequency, the decoherence of the different spin waves right under the emitting antenna happens short after the time origin, and nothing but noise and artefactual base line is detected in the impulse responses $s_{11}(t)$ and $s_{22}(t)$ after the few first 100 ps. The contributions of the different spin waves thus cannot be separated in the reflection signals, and $\tilde{S}_{11}$ and $\tilde{S}_{22}$ cannot serve our present objectives, except for a crude estimation of the antenna efficiency function \cite{sushruth_electrical_2020}.  

%
\begin{figure}
\hspace*{-0.6cm}\includegraphics[width=9 cm]{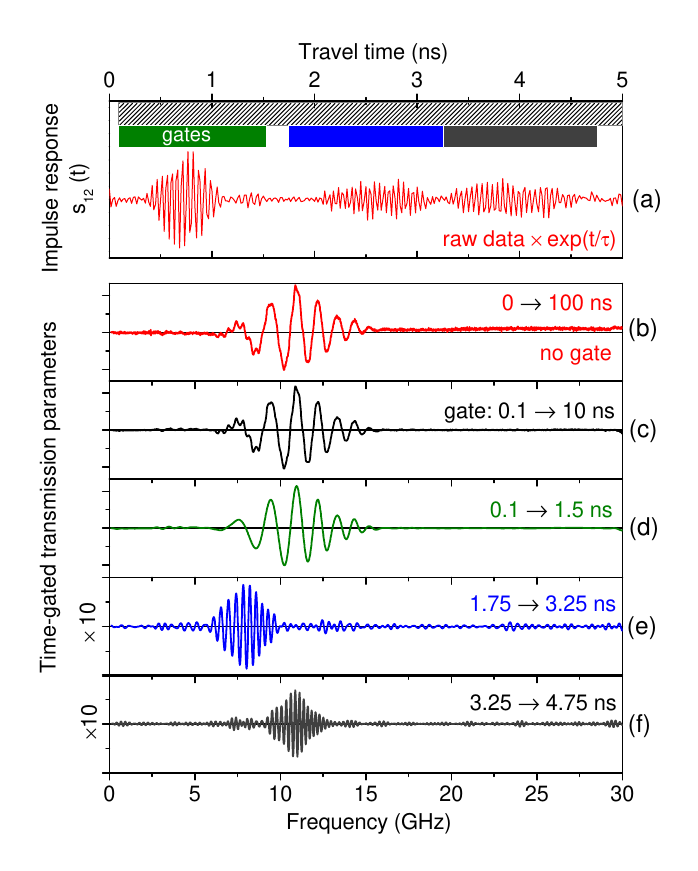}{\centering}
\caption{Time gating method to isolate the spectra of single modes in experimental data. (a) Exponentially amplified impulse response ${s}_{21}(t) e^{t / \tau}$ for spin-wave conduits of width of $4.7~\mu\textrm{m}$ at 18 mT after a travel distance of $r_3=6.9~\mu\textrm{m}$ corresponding to the raw spectrum in (b). The attenuation rate is taken as that of FMR, i.e. $\tau=2/ \Delta f$ . (c-f): real parts of the time-gated $\tilde{S}_{12}$ spectra. In (c), the feed-through and the long travel time were removed, effectively suppressing the base line and the high frequency noise. In (d), the fastest wavepacket is kept to isolate the contribution of a single mode.  In (e), the medium velocity wavepacket is kept. In (d), the slowest wavepacket is kept. The vertical scales in (c) and (d) were multiplied by 10. }
\label{fig4}
\end{figure}


\section{Construction of the dispersion relation of each family of spin waves using time-gating} \label{constructionOfDispersion}
We now aim to isolate the contribution of each family of spin wave modes in the device response in order to later deduce the dispersion relation of each family separately. The figure~\ref{fig4} illustrates this procedure for the widest spin-wave conduit.
\subsection{Time-gating for the selection of the contribution of a single family of spin waves}
This is performed by time-gating the impulse response and transforming back the data to frequency domain. The procedure requires that the wavepackets are separated in time. For two dispersionless modes of group velocities $v_{g1}$ and $v_{g2}$ and same linewidth $\Delta f$, this time-separation is ensured for long enough propagation distances, when 
\begin{equation}   \Big| \frac{r}{ v_{g1}}-\frac{r}{ v_{g2}} \Big| \geq \frac{2}{\Delta f }.\label{CriterionWavepacketSeparation}\end{equation} This condition is clearly satisfied in our cases (Fig.~\ref{fig3}), so we can use abrupt gates consisting of rectangular windows starting at $t_\textrm{start}$ and ending at $t_\textrm{end}$. The gated spectra are defined as:
\begin{equation} 
\tilde{S}_{ij}^\textrm{gated} =  {\mathscr F^{-1}} \Big[ {\mathscr F} [ \tilde{S}_{ij}^\textrm{corr} (f) ] .\Theta (t-t_\textrm{start}) . \Theta(t_\textrm{end}-t) \Big],
\label{gating} 
\end{equation}
where $\Theta$ is the Heaviside function. This procedure is illustrated in Fig.~\ref{fig4} for the widest spin wave conduit.
Although this is not of direct use for our present purpose, we would like to mention that it is possible to get better-looking data in a rigorous manner illustrated in Fig.~\ref{fig4}(b, c). The (non-spin-wave-related) slowly-varying base line in the $\tilde{S}_{ij}^\textrm{gated}$ spectra is effectively removed by gating out the residue of the feed-through by choosing $t_\textrm{start}=100~\textrm{ps}$. In addition, the trace noise in $\tilde{S}_{ij}^\textrm{gated}$ is much reduced by setting for instance $t_\textrm{end}=10~\textrm{ns}$, i.e. by gating out the noise-dominated signals that arrive at the receiving antenna late after the spin waves. 

Let us now select single families of spin wave modes. To select the fastest branch, we gate out both the feed-through signal as well as the late-arriving wavepackets, as shown in Fig.~\ref{fig4}(d). The transformed data looks now very similar to the signal expected in the single mode theory [Eq.~\ref{1DmonomodeTheory} and Fig.~\ref{fig2}(e)]. The contributions of the medium velocity and low velocity families of modes can also be constructed [see Fig.~\ref{fig4}(e, f)]; they also resemble the signal expected in the single mode theory but with a much weaker amplitude. We can thus try and use Eq.~\ref{1DmonomodeTheory} to derive the dispersion relation $\omega(k_x)$ of each family of spin waves.

%
\begin{figure}
\hspace*{-0.1cm}\includegraphics[width=8.5 cm]{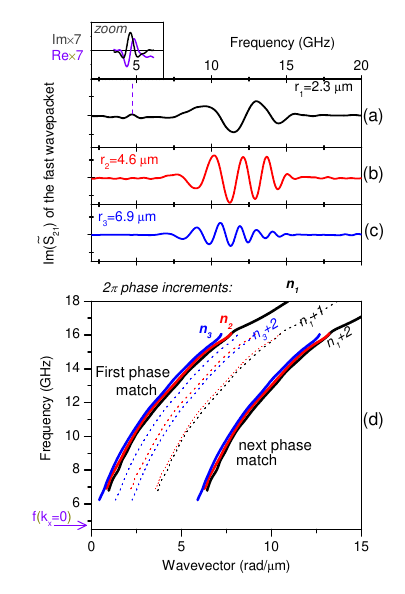}{\centering}
\caption{Method to determine the absolute phase of the transmitted spin wave signals and to disambiguate the wavevectors. (a, b, c): imaginary parts of $\tilde{S}_{12}$ for propagation distances $r_1$, $r_2$ and $r_3$ at 18 mT for $w_\textrm{mag}=4.7~\mu \textrm{m}$. The time gating was set to keep the feed-through and the first arriving wavepacket. 
Zoom within panel (a): complex signal related to the distant excitation of the $k_x=0$ mode. Panel (d): Dispersion relations obtained from Eq.~\ref{k} for each propagation distance ($r_1$: black; $r_2$: red; $r_3$: blue). The bold curves that almost superimpose correspond to the mathematically correct choices of the integration constants ${n_1, n_2, n_3}$ within Eq.~\ref{k}.. Only the first phase match is both physically and mathematically correct. The dotted lines are when errors of $2\pi$ and $4\pi$ are done within Eq.~\ref{k}.}
\label{fig5}
\end{figure}

\subsection{Transforming scattering parameters in dispersion relations}
\subsubsection{Determination of the frequency dependence of the group velocity}
In the single mode theory (Eq.~\ref{1DmonomodeTheory}), the phase of the transmission parameters solely depends on the propagation distance $r$ and the wavevector $k_x$. This phase can be evaluated from the experimental data by noticing that: 
\begin{equation}
\frac{\tilde{S}_{ij}}{ || \tilde{S}_{ij} ||} = e^{-i k_x |r|} \label{phase}
\end{equation}
such that the wavevector at a given frequency can be found by unwrapping the phase:
\begin{equation}
k_x(\omega) = - \frac{1}{r} \big[\textrm{Arg} \frac{\tilde{S}_{ij}(\omega)}{ || \tilde{S}_{ij}(\omega) ||}\big] + \frac{2n \pi}{r},~n \in \mathbb{N}
\label{k}
\end{equation}
Because of the $2\pi$ indetermination of the absolute phase, the above expression is only sufficient to calculate the group velocity $\frac{\partial \omega}{\partial k_x}$ versus frequency, but not sufficient to define the wavevector in a unique manner. This difficulty can be solved by the determination of the absolute phase of Eq.~\ref{phase}. 
\subsubsection{Unequivocal determination of the wavevectors: from the uniform modes} 
The most straightforward method would be to measure independently one point of each dispersion curve. The $k_x=0$ point of each dispersion curve is the natural choice. Unfortunately measuring the $k_x=0$ point of each dispersion is difficult in a \textit{propagation} experiment as this would require to (i) be able to excite at $k_x=0$ and (ii) to ensure that the $k_x=0$ reaches the receiving antenna. The first condition is trivial to obtain by the use of single-wire antennas \cite{ciubotaru_all_2016, qin_propagating_2018}. 
However as mentioned in section \ref{design} the second condition is generally not met if the spin wave conduit has a finite width, since in this case the $k_x=0$ modes have vanishing group velocities and never reach the receiving antenna. We thus measure the $k_x=0$ point in a different manner.

The measurement of the $k_x=0$ frequency can sometimes be done thanks to a collateral effect called the \textit{distant} induction. Indeed part of the magnetization waves existing below the receiving antenna can be directly inductively excited there by the long-range rf field produced by the emitting antenna \cite{birt_deviation_2012}. This distant-induction signal starts at a time $r/c_g^\textrm{em}$, therefore it disappears if time-gating is set to exclude the signals arriving immediately at the receiving antenna. As the long-range rf field of the emitting antenna varies slowly in space, the shape of the distant induction signal should resemble the susceptibility $\chi(\omega, k_x=0)$ of the excited mode (see Eq. 16 in ref.~\cite{sushruth_electrical_2020}). 
In the transmission coefficient, we can sometimes identify this feature, as for instance in Fig.~\ref{fig5}(a), see the zoom at 4.7 GHz. These tiny symmetric/asymmetric Lorentzian features in the imaginary/real parts of the signal are arguments to state that $f(k_x=0)=4.7~\textrm{GHz}$ for at east one mode of the widest conduit. The linewidth of these features at 4.7 GHz is $220\pm20~\textrm{MHz}$, i.e. very similar to that of the FMR of the unpatterned film at the same frequency. Unfortunately, the signal-to-noise ratio for the other modes is not always high enough to perceive the $k_x=0$ points, such that a more general procedure is needed.

\subsubsection{Unequivocal determination of the wavevectors: from the several propagation distances} 
We have thus implemented another method that solely relies on spin wave propagation and therefore works for all detectable modes. This procedure is illustrated in Fig.~\ref{fig5} for the fastest mode of the widest conduit. The idea is to apply Eq.~\ref{k} for several propagation distances $\{r_1,~r_2,...\}$ and find the corresponding sets of integers $\{n_1,~n_2,...\}$  that make the results of Eq. ~\ref{k} mathematically match. In principle, if two incommensurate propagation distances are used, there is a unique pair $\{n_1,~n_2\}$ that enables the mathematical matching. If the propagation distances are commensurate, there is an infinite number of sets of integers $\{n_1,~n_2,...\}$ that satisfy Eq. ~\ref{k}. The physically relevant one must be found in these sets.  \\We are in this situation since our propagation distances are commensurate and obey $r_2=2 r_1$ and $r_3=3 r_1$. We thus first identify the triplets $\{n_1,~n_2,~n_3\}$  so that $k_x r_2 = 2 k_x r_1$ and $k_x r_3 = 3 k_x r_1$ to get the mathematical matching of the three Eq.~\ref{k}, as illustrated in Fig.~\ref{fig5}(d). This is done by starting with small values of $n_1$ (say 0, 1, 2 or 3) at the lowest frequency at which the propagating spin wave is detected.  Looking at Fig. 5, we can assess that only the first working triplet $\{n_1,~n_2,~n_3\}$ leads to a physically possible dispersion relation: the next triplets that are mathematically correct and lead to phase matching clearly generate non-physical dispersion relations that can be discarded.
The dispersion curve deduced from the first triplet seems to extrapolate to the point $f(k_x=0)=4.7~\textrm{GHz}$ formerly determined by the distant induction. This consistency clearly supports our procedure.

\subsection{Discussion on the nature of each family of modes} 
The figures 5d and 6a display the dispersion relation of the fastest mode of the widest conduit. As anticipated from the large width of the corresponding wavepacket in time domain, this mode is clearly dispersive: within this branch, the spin waves with different wavevectors/frequencies travel at different group velocities such that the wavepacket spreads as it travels. The group velocity decreases from 14 km/s at $k_x\approx 1$ to 5 km/s at 11 rad/$\mu$m. The large amplitude of this mode argues for the assignment of this mode to the DE$_1$ mode. To further support this assertion, we have plotted on Fig.~\ref{fig6}(a) the theoretical dispersion curve of this mode (Eq.~\ref{DE}). The agreement is perfect with no fitting parameters, provided that the internal field $H_0$ is taken as the applied field $H_y$ minus the dipolar shape anisotropy field.

Although the agreement between the experimental dispersion curves and the theoretical ones is less satisfactory, the two fastest mode of the narrowest conduit can be faithfully assigned to the DE$_1$ and DE$_3$ modes. The difference stems probably \cite{talmelli_electrical_2021} from an imperfect saturation of the magnetization in the experiments, while the model assumes perfect saturation. In contrast, such an assignment cannot be done for the two slowest mode of the widest conduit. We believe that they are edge modes, but accounting exactly for these dispersion curves is beyond the scope of the present study. Another argument supporting this interpretation is the fact that these modes progressively disappear when increasing the applied field far above the saturation field (not shown). 

\subsection{Consistency check and revisit of the antenna efficiency function}
The self consistency of our approach can be finally checked by using the modulus of the transmission parameters for a single spin wave branch $|| \tilde{S}_{ij}^{\textrm{gated}}(\omega) ||$ and by plotting it versus the determined $k_x$ using the experimental dispersion relation $\omega(k_x)$. From the single mode theory, this modulus should be proportional to $e^{-\frac{|r|}{L_\textrm{att}}} ~(h_{x}(k_x))^2$ and thus $|| \tilde{S}_{ij}^{\textrm{gated}}(k_x) ||$ should ressemble the idealized antenna efficiency function. The zeros of the two functions at $k_x^\textrm{node} \in \{0, \frac{2\pi}{g+L}, \frac{2\pi}{L},.. \}$ should match perfectly. The maxima at $ k_x^\textrm{antinode}$ should match perfectly only for the non-dispersive modes because otherwise $L_\textrm{att}$ is a function of $k_x$ . 

\begin{figure}
\hspace*{-0.5cm}\includegraphics[width=9.5 cm]{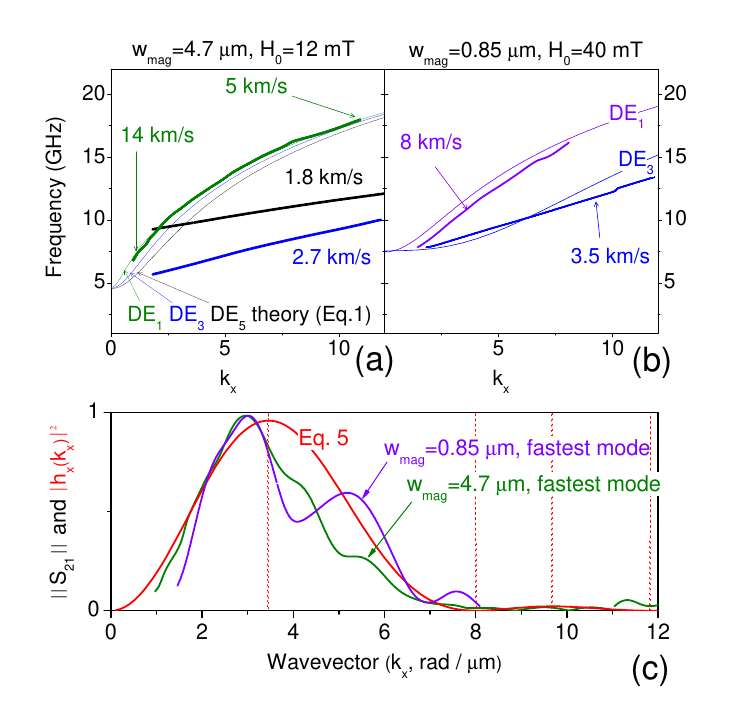}{\centering}
\caption{(a) and (b): Experimental dispersion relations (bold lines) for spin wave conduits of widths 4.7 and $0.85~\mu\textrm{m}$. Narrow lines: theoretical dispersion relations (Eq.~\ref{DE}) of the DE$_1$, DE$_2$ and DE$_3$ modes. In the models, the internal fields $H_0$ are taken as the experimental applied fields (18 and 70 mT) minus the experimental shape anisotropy fields (6 and 30 mT) of the conduits.  (c): Comparison between the antenna efficiency function (Eq.~\ref{AntennaEfficiencyFunction}) and the modulus of the transmission coefficients of the fastest mode for $r_3=6.9~\mu\textrm{m}$. Green curve: for $w=4.7~\mu\textrm{m}$ and $\mu_0 H_y=18~\textrm{mT}$.  Violet curve: for $w=0.85~\mu\textrm{m}$ and $\mu_0 H_y=70~\textrm{mT}$. The vertical dotted lines indicate the theoretical positions of the zeros and the maxima of $h_x(k_x)^2$. }
\label{fig6}
\end{figure}

This consistency check is done in Fig.~\ref{fig6}(c) for the fastest modes of the two conduits, i.e. the DE$_1$'s. The maximum efficiency of the antenna, expected at $k_x^\textrm{antinode}=3.5~\textrm{rad/}\mu\textrm{m}$ is found experimentally at $3.0~\textrm{rad/}\mu\textrm{m}$ instead. The zeros of $|| \tilde{S}_{ij}^{\textrm{gated}}(k_x) ||$ are hard to identify exactly because they are obscured by the trace noise ; however, the experimental $k_\textrm{node}$ are found to match poorly with the expectations from Eq.~\ref{AntennaEfficiencyFunction}. This questions the validity of the past studies in which the spin wave wavevectors were postulated from the expected efficiency function of an idealized antenna. Understanding this difference between $|| \tilde{S}_{ij}^{\textrm{gated}}(k_x) ||$ and Eq.~\ref{AntennaEfficiencyFunction} would require to calculate the profile of the rf fields generated by the antenna in the presence of the magnetic medium with its magnetic susceptibility and its electrical conductance in an exact manner. This task is beyond the scope of the present paper.

%

\section{Summary}
In summary, we have developed time-of-flight spin wave spectroscopy to map the band structure of the spin waves in a nanostructured  film. Our method is based on the scattering matrix of a network of inductive antennas emitting and collecting the spin waves (Fig.~\ref{fig1}). As the signal analysis is based on the phase of the spin waves after propagation, specific experimental precautions must be taken while designing the length of the microwave circuitry (Eq.~\ref {eqell}). \\
In general, several spin wave branches contribute to this signal which renders this signal complicated to account for [Fig.~\ref{fig2}(d)]. The construction of the dispersion relation of a given spin wave branch requires to isolate its contribution. If the spin wave branches have different group velocities they can be conveniently sorted by time-of-flight spectroscopy. For this purpose, we mathematically transform the data to get the transmission impulse response in time-domain using Eq.~\ref{FFT}. The different spin wave branches are viewed as wavepackets that reach successively the receiving antenna after different travel times (Fig.~\ref{fig3}). 
The wavepackets can then be separated by time-gating (Eq.~\ref{gating}) if a large enough propagation distance (Eq.~\ref{CriterionWavepacketSeparation}) is used.

The time-gated responses are then used to calculate the contribution of a each spin wave branch to the frequency domain scattering matrix, which recovers an intuitive shape (Fig.~\ref{fig4}). Under reasonable assumptions \cite{sushruth_electrical_2020}, the dispersion relation of a branch of spin waves stems from the absolute phase of the transmission parameter related to this branch (Eq.~\ref{1DmonomodeTheory}). Unfortunately the phase can only be determined (Eq.~\ref{k}) with a $2 \pi$ uncertainty. This difficulty can be circumvented if several propagation distances are harnessed (Fig.~\ref{fig5}). Alternatively, it can be solved by an independent measurement of one point of the dispersion curve, for instance the $k_x=0$ one. 
Finally, the consistency of the whole procedure can be checked (Fig.~\ref{fig6}) by the back calculation of the antenna emission spectrum in reciprocal space and, when possible, by a comparison to the expected spin wave band structure. 
\textcolor{black}{Our method should be applicable to various situations, and in particular it should be well adapted to spin-waves that are within the exchange regime \cite{chen_excitation_2019} where other characterization techniques like Brillouin Light Scattering are difficult to implement.}

\section{Acknowledgment}
This work was funded through the FETOPEN-01-2016-2017—FET-Open research and innovation actions (CHIRON project: Grant Agreement No. 801055) and was partly supported by the French RENATECH network. This work is also supported by a public grant overseen by the French National Research Agency (ANR) as part of the “Investissements d’Avenir” program (Labex NanoSaclay, reference: ANR-10-LABX-0035). Contributions:  G.T. fabricated the samples described in this paper, under the supervision of F.C and C.A.. S.M.N. fabricated other samples on which the methodology was developed. G.T. and T.D. made the experiments. T.D. developed the set-up, the analytics and made the data analysis. C.C. and T.D. supervised the study. T.D. thanks Umesh K. Bhaskar for useful discussions.


\begin{thebibliography}{30}%
\makeatletter
\providecommand \@ifxundefined [1]{%
 \@ifx{#1\undefined}
}%
\providecommand \@ifnum [1]{%
 \ifnum #1\expandafter \@firstoftwo
 \else \expandafter \@secondoftwo
 \fi
}%
\providecommand \@ifx [1]{%
 \ifx #1\expandafter \@firstoftwo
 \else \expandafter \@secondoftwo
 \fi
}%
\providecommand \natexlab [1]{#1}%
\providecommand \enquote  [1]{``#1''}%
\providecommand \bibnamefont  [1]{#1}%
\providecommand \bibfnamefont [1]{#1}%
\providecommand \citenamefont [1]{#1}%
\providecommand \href@noop [0]{\@secondoftwo}%
\providecommand \href [0]{\begingroup \@sanitize@url \@href}%
\providecommand \@href[1]{\@@startlink{#1}\@@href}%
\providecommand \@@href[1]{\endgroup#1\@@endlink}%
\providecommand \@sanitize@url [0]{\catcode `\\12\catcode `\$12\catcode
  `\&12\catcode `\#12\catcode `\^12\catcode `\_12\catcode `\%12\relax}%
\providecommand \@@startlink[1]{}%
\providecommand \@@endlink[0]{}%
\providecommand \url  [0]{\begingroup\@sanitize@url \@url }%
\providecommand \@url [1]{\endgroup\@href {#1}{\urlprefix }}%
\providecommand \urlprefix  [0]{URL }%
\providecommand \Eprint [0]{\href }%
\providecommand \doibase [0]{http://dx.doi.org/}%
\providecommand \selectlanguage [0]{\@gobble}%
\providecommand \bibinfo  [0]{\@secondoftwo}%
\providecommand \bibfield  [0]{\@secondoftwo}%
\providecommand \translation [1]{[#1]}%
\providecommand \BibitemOpen [0]{}%
\providecommand \bibitemStop [0]{}%
\providecommand \bibitemNoStop [0]{.\EOS\space}%
\providecommand \EOS [0]{\spacefactor3000\relax}%
\providecommand \BibitemShut  [1]{\csname bibitem#1\endcsname}%
\let\auto@bib@innerbib\@empty
\bibitem [{\citenamefont {Kruglyak}, \citenamefont {Demokritov},\ and\
  \citenamefont {Grundler}(2010)}]{Kruglyak_magnonics_2010}%
  \BibitemOpen
  \bibfield  {author} {\bibinfo {author} {\bibfnamefont {V.~V.}\ \bibnamefont
  {Kruglyak}}, \bibinfo {author} {\bibfnamefont {S.~O.}\ \bibnamefont
  {Demokritov}}, \ and\ \bibinfo {author} {\bibfnamefont {D.}~\bibnamefont
  {Grundler}},\ }\href {\doibase 10.1088/0022-3727/43/26/264001} {\bibfield
  {journal} {\bibinfo  {journal} {Journal of Physics D: Applied Physics}\
  }\textbf {\bibinfo {volume} {43}},\ \bibinfo {pages} {264001} (\bibinfo
  {year} {2010})},\ \bibinfo {note} {publisher: IOP Publishing}\BibitemShut
  {NoStop}%
\bibitem [{\citenamefont {Chumak}\ \emph {et~al.}(2015)\citenamefont {Chumak},
  \citenamefont {Vasyuchka}, \citenamefont {Serga},\ and\ \citenamefont
  {Hillebrands}}]{chumak_magnon_2015}%
  \BibitemOpen
  \bibfield  {author} {\bibinfo {author} {\bibfnamefont {A.~V.}\ \bibnamefont
  {Chumak}}, \bibinfo {author} {\bibfnamefont {V.~I.}\ \bibnamefont
  {Vasyuchka}}, \bibinfo {author} {\bibfnamefont {A.~A.}\ \bibnamefont
  {Serga}}, \ and\ \bibinfo {author} {\bibfnamefont {B.}~\bibnamefont
  {Hillebrands}},\ }\href {\doibase 10.1038/nphys3347} {\bibfield  {journal}
  {\bibinfo  {journal} {Nature Physics}\ }\textbf {\bibinfo {volume} {11}},\
  \bibinfo {pages} {453} (\bibinfo {year} {2015})}\BibitemShut {NoStop}%
\bibitem [{\citenamefont {Hillebrands}(1999)}]{hillebrands_progress_1999}%
  \BibitemOpen
  \bibfield  {author} {\bibinfo {author} {\bibfnamefont {B.}~\bibnamefont
  {Hillebrands}},\ }\href {\doibase 10.1063/1.1149637} {\bibfield  {journal}
  {\bibinfo  {journal} {Review of Scientific Instruments}\ }\textbf {\bibinfo
  {volume} {70}},\ \bibinfo {pages} {1589} (\bibinfo {year} {1999})},\ \bibinfo
  {note} {publisher: American Institute of Physics}\BibitemShut {NoStop}%
\bibitem [{\citenamefont {Demokritov}\ and\ \citenamefont
  {Demidov}(2008)}]{demokritov_micro-brillouin_2008}%
  \BibitemOpen
  \bibfield  {author} {\bibinfo {author} {\bibfnamefont {S.~O.}\ \bibnamefont
  {Demokritov}}\ and\ \bibinfo {author} {\bibfnamefont {V.~E.}\ \bibnamefont
  {Demidov}},\ }\href {\doibase 10.1109/TMAG.2007.910227} {\bibfield  {journal}
  {\bibinfo  {journal} {IEEE Transactions on Magnetics}\ }\textbf {\bibinfo
  {volume} {44}},\ \bibinfo {pages} {6} (\bibinfo {year} {2008})},\ \bibinfo
  {note} {conference Name: IEEE Transactions on Magnetics}\BibitemShut
  {NoStop}%
\bibitem [{\citenamefont {Sebastian}\ \emph {et~al.}(2015)\citenamefont
  {Sebastian}, \citenamefont {Schultheiss}, \citenamefont {Obry}, \citenamefont
  {Hillebrands},\ and\ \citenamefont
  {Schultheiss}}]{sebastian_micro-focused_2015}%
  \BibitemOpen
  \bibfield  {author} {\bibinfo {author} {\bibfnamefont {T.}~\bibnamefont
  {Sebastian}}, \bibinfo {author} {\bibfnamefont {K.}~\bibnamefont
  {Schultheiss}}, \bibinfo {author} {\bibfnamefont {B.}~\bibnamefont {Obry}},
  \bibinfo {author} {\bibfnamefont {B.}~\bibnamefont {Hillebrands}}, \ and\
  \bibinfo {author} {\bibfnamefont {H.}~\bibnamefont {Schultheiss}},\ }\href
  {\doibase 10.3389/fphy.2015.00035} {\bibfield  {journal} {\bibinfo  {journal}
  {Frontiers in Physics}\ }\textbf {\bibinfo {volume} {3}} (\bibinfo {year}
  {2015}),\ 10.3389/fphy.2015.00035},\ \bibinfo {note} {publisher:
  Frontiers}\BibitemShut {NoStop}%
\bibitem [{\citenamefont {Bailleul}, \citenamefont {Olligs},\ and\
  \citenamefont {Fermon}(2003)}]{bailleul_propagating_2003}%
  \BibitemOpen
  \bibfield  {author} {\bibinfo {author} {\bibfnamefont {M.}~\bibnamefont
  {Bailleul}}, \bibinfo {author} {\bibfnamefont {D.}~\bibnamefont {Olligs}}, \
  and\ \bibinfo {author} {\bibfnamefont {C.}~\bibnamefont {Fermon}},\ }\href
  {\doibase 10.1063/1.1597745} {\bibfield  {journal} {\bibinfo  {journal}
  {Applied Physics Letters}\ }\textbf {\bibinfo {volume} {83}},\ \bibinfo
  {pages} {972} (\bibinfo {year} {2003})},\ \bibinfo {note} {publisher:
  American Institute of Physics}\BibitemShut {NoStop}%
\bibitem [{\citenamefont {Ciubotaru}\ \emph {et~al.}(2016)\citenamefont
  {Ciubotaru}, \citenamefont {Devolder}, \citenamefont {Manfrini},
  \citenamefont {Adelmann},\ and\ \citenamefont {Radu}}]{ciubotaru_all_2016}%
  \BibitemOpen
  \bibfield  {author} {\bibinfo {author} {\bibfnamefont {F.}~\bibnamefont
  {Ciubotaru}}, \bibinfo {author} {\bibfnamefont {T.}~\bibnamefont {Devolder}},
  \bibinfo {author} {\bibfnamefont {M.}~\bibnamefont {Manfrini}}, \bibinfo
  {author} {\bibfnamefont {C.}~\bibnamefont {Adelmann}}, \ and\ \bibinfo
  {author} {\bibfnamefont {I.~P.}\ \bibnamefont {Radu}},\ }\href {\doibase
  10.1063/1.4955030} {\bibfield  {journal} {\bibinfo  {journal} {Applied
  Physics Letters}\ }\textbf {\bibinfo {volume} {109}},\ \bibinfo {pages}
  {012403} (\bibinfo {year} {2016})}\BibitemShut {NoStop}%
\bibitem [{\citenamefont {Talmelli}\ \emph {et~al.}(2021)\citenamefont
  {Talmelli}, \citenamefont {Narducci}, \citenamefont {Vanderveken},
  \citenamefont {Heyns}, \citenamefont {Irrera}, \citenamefont {Asselberghs},
  \citenamefont {Radu}, \citenamefont {Adelmann},\ and\ \citenamefont
  {Ciubotaru}}]{talmelli_electrical_2021}%
  \BibitemOpen
  \bibfield  {author} {\bibinfo {author} {\bibfnamefont {G.}~\bibnamefont
  {Talmelli}}, \bibinfo {author} {\bibfnamefont {D.}~\bibnamefont {Narducci}},
  \bibinfo {author} {\bibfnamefont {F.}~\bibnamefont {Vanderveken}}, \bibinfo
  {author} {\bibfnamefont {M.}~\bibnamefont {Heyns}}, \bibinfo {author}
  {\bibfnamefont {F.}~\bibnamefont {Irrera}}, \bibinfo {author} {\bibfnamefont
  {I.}~\bibnamefont {Asselberghs}}, \bibinfo {author} {\bibfnamefont {I.~P.}\
  \bibnamefont {Radu}}, \bibinfo {author} {\bibfnamefont {C.}~\bibnamefont
  {Adelmann}}, \ and\ \bibinfo {author} {\bibfnamefont {F.}~\bibnamefont
  {Ciubotaru}},\ }\href {http://arxiv.org/abs/2101.11983} {\bibfield  {journal}
  {\bibinfo  {journal} {arXiv:2101.11983 [cond-mat, physics:physics]}\ }
  (\bibinfo {year} {2021})},\ \bibinfo {note} {arXiv: 2101.11983}\BibitemShut
  {NoStop}%
\bibitem [{\citenamefont {Yu}\ \emph {et~al.}(2012)\citenamefont {Yu},
  \citenamefont {Huber}, \citenamefont {Schwarze}, \citenamefont {Brandl},
  \citenamefont {Rapp}, \citenamefont {Berberich}, \citenamefont {Duerr},\ and\
  \citenamefont {Grundler}}]{yu_high_2012}%
  \BibitemOpen
  \bibfield  {author} {\bibinfo {author} {\bibfnamefont {H.}~\bibnamefont
  {Yu}}, \bibinfo {author} {\bibfnamefont {R.}~\bibnamefont {Huber}}, \bibinfo
  {author} {\bibfnamefont {T.}~\bibnamefont {Schwarze}}, \bibinfo {author}
  {\bibfnamefont {F.}~\bibnamefont {Brandl}}, \bibinfo {author} {\bibfnamefont
  {T.}~\bibnamefont {Rapp}}, \bibinfo {author} {\bibfnamefont {P.}~\bibnamefont
  {Berberich}}, \bibinfo {author} {\bibfnamefont {G.}~\bibnamefont {Duerr}}, \
  and\ \bibinfo {author} {\bibfnamefont {D.}~\bibnamefont {Grundler}},\ }\href
  {\doibase 10.1063/1.4731273} {\bibfield  {journal} {\bibinfo  {journal}
  {Applied Physics Letters}\ }\textbf {\bibinfo {volume} {100}},\ \bibinfo
  {pages} {262412} (\bibinfo {year} {2012})}\BibitemShut {NoStop}%
\bibitem [{\citenamefont {Maendl}, \citenamefont {Stasinopoulos},\ and\
  \citenamefont {Grundler}(2017)}]{maendl_spin_2017}%
  \BibitemOpen
  \bibfield  {author} {\bibinfo {author} {\bibfnamefont {S.}~\bibnamefont
  {Maendl}}, \bibinfo {author} {\bibfnamefont {I.}~\bibnamefont
  {Stasinopoulos}}, \ and\ \bibinfo {author} {\bibfnamefont {D.}~\bibnamefont
  {Grundler}},\ }\href {\doibase 10.1063/1.4991520} {\bibfield  {journal}
  {\bibinfo  {journal} {Applied Physics Letters}\ }\textbf {\bibinfo {volume}
  {111}},\ \bibinfo {pages} {012403} (\bibinfo {year} {2017})},\ \bibinfo
  {note} {publisher: American Institute of Physics}\BibitemShut {NoStop}%
\bibitem [{\citenamefont {Qin}\ \emph {et~al.}(2018)\citenamefont {Qin},
  \citenamefont {Hämäläinen}, \citenamefont {Arjas}, \citenamefont
  {Witteveen},\ and\ \citenamefont {van Dijken}}]{qin_propagating_2018}%
  \BibitemOpen
  \bibfield  {author} {\bibinfo {author} {\bibfnamefont {H.}~\bibnamefont
  {Qin}}, \bibinfo {author} {\bibfnamefont {S.~J.}\ \bibnamefont
  {Hämäläinen}}, \bibinfo {author} {\bibfnamefont {K.}~\bibnamefont
  {Arjas}}, \bibinfo {author} {\bibfnamefont {J.}~\bibnamefont {Witteveen}}, \
  and\ \bibinfo {author} {\bibfnamefont {S.}~\bibnamefont {van Dijken}},\
  }\href {\doibase 10.1103/PhysRevB.98.224422} {\bibfield  {journal} {\bibinfo
  {journal} {Physical Review B}\ }\textbf {\bibinfo {volume} {98}},\ \bibinfo
  {pages} {224422} (\bibinfo {year} {2018})},\ \bibinfo {note} {publisher:
  American Physical Society}\BibitemShut {NoStop}%
\bibitem [{\citenamefont {Stückler}\ \emph {et~al.}(2017)\citenamefont
  {Stückler}, \citenamefont {Liu}, \citenamefont {Liu}, \citenamefont {Yu},
  \citenamefont {Heimbach}, \citenamefont {Chen}, \citenamefont {Hu},
  \citenamefont {Tu}, \citenamefont {Zhang}, \citenamefont {Granville},
  \citenamefont {Wu}, \citenamefont {Liao}, \citenamefont {Yu},\ and\
  \citenamefont {Zhao}}]{Stuckler_ultrabroadband_2017}%
  \BibitemOpen
  \bibfield  {author} {\bibinfo {author} {\bibfnamefont {T.}~\bibnamefont
  {Stückler}}, \bibinfo {author} {\bibfnamefont {C.}~\bibnamefont {Liu}},
  \bibinfo {author} {\bibfnamefont {T.}~\bibnamefont {Liu}}, \bibinfo {author}
  {\bibfnamefont {H.}~\bibnamefont {Yu}}, \bibinfo {author} {\bibfnamefont
  {F.}~\bibnamefont {Heimbach}}, \bibinfo {author} {\bibfnamefont
  {J.}~\bibnamefont {Chen}}, \bibinfo {author} {\bibfnamefont {J.}~\bibnamefont
  {Hu}}, \bibinfo {author} {\bibfnamefont {S.}~\bibnamefont {Tu}}, \bibinfo
  {author} {\bibfnamefont {Y.}~\bibnamefont {Zhang}}, \bibinfo {author}
  {\bibfnamefont {S.}~\bibnamefont {Granville}}, \bibinfo {author}
  {\bibfnamefont {M.}~\bibnamefont {Wu}}, \bibinfo {author} {\bibfnamefont
  {Z.-M.}\ \bibnamefont {Liao}}, \bibinfo {author} {\bibfnamefont
  {D.}~\bibnamefont {Yu}}, \ and\ \bibinfo {author} {\bibfnamefont
  {W.}~\bibnamefont {Zhao}},\ }\href {\doibase 10.1103/PhysRevB.96.144430}
  {\bibfield  {journal} {\bibinfo  {journal} {Physical Review B}\ }\textbf
  {\bibinfo {volume} {96}},\ \bibinfo {pages} {144430} (\bibinfo {year}
  {2017})},\ \bibinfo {note} {publisher: American Physical Society}\BibitemShut
  {NoStop}%
\bibitem [{\citenamefont {Chen}\ \emph {et~al.}(2018)\citenamefont {Chen},
  \citenamefont {Heimbach}, \citenamefont {Liu}, \citenamefont {Yu},
  \citenamefont {Liu}, \citenamefont {Chang}, \citenamefont {Stückler},
  \citenamefont {Hu}, \citenamefont {Zeng}, \citenamefont {Zhang},
  \citenamefont {Liao}, \citenamefont {Yu}, \citenamefont {Zhao},\ and\
  \citenamefont {Wu}}]{chen_spin_2018}%
  \BibitemOpen
  \bibfield  {author} {\bibinfo {author} {\bibfnamefont {J.}~\bibnamefont
  {Chen}}, \bibinfo {author} {\bibfnamefont {F.}~\bibnamefont {Heimbach}},
  \bibinfo {author} {\bibfnamefont {T.}~\bibnamefont {Liu}}, \bibinfo {author}
  {\bibfnamefont {H.}~\bibnamefont {Yu}}, \bibinfo {author} {\bibfnamefont
  {C.}~\bibnamefont {Liu}}, \bibinfo {author} {\bibfnamefont {H.}~\bibnamefont
  {Chang}}, \bibinfo {author} {\bibfnamefont {T.}~\bibnamefont {Stückler}},
  \bibinfo {author} {\bibfnamefont {J.}~\bibnamefont {Hu}}, \bibinfo {author}
  {\bibfnamefont {L.}~\bibnamefont {Zeng}}, \bibinfo {author} {\bibfnamefont
  {Y.}~\bibnamefont {Zhang}}, \bibinfo {author} {\bibfnamefont
  {Z.}~\bibnamefont {Liao}}, \bibinfo {author} {\bibfnamefont {D.}~\bibnamefont
  {Yu}}, \bibinfo {author} {\bibfnamefont {W.}~\bibnamefont {Zhao}}, \ and\
  \bibinfo {author} {\bibfnamefont {M.}~\bibnamefont {Wu}},\ }\href {\doibase
  10.1016/j.jmmm.2017.04.045} {\bibfield  {journal} {\bibinfo  {journal}
  {Journal of Magnetism and Magnetic Materials}\ }\bibinfo {series}
  {Perspectives on magnon spintronics},\ \textbf {\bibinfo {volume} {450}},\
  \bibinfo {pages} {3} (\bibinfo {year} {2018})}\BibitemShut {NoStop}%
\bibitem [{\citenamefont {Chen}\ \emph
  {et~al.}(2019{\natexlab{a}})\citenamefont {Chen}, \citenamefont {Wang},
  \citenamefont {Liu}, \citenamefont {Tu}, \citenamefont {Bi},\ and\
  \citenamefont {Yu}}]{chen_spin_2019}%
  \BibitemOpen
  \bibfield  {author} {\bibinfo {author} {\bibfnamefont {J.}~\bibnamefont
  {Chen}}, \bibinfo {author} {\bibfnamefont {C.}~\bibnamefont {Wang}}, \bibinfo
  {author} {\bibfnamefont {C.}~\bibnamefont {Liu}}, \bibinfo {author}
  {\bibfnamefont {S.}~\bibnamefont {Tu}}, \bibinfo {author} {\bibfnamefont
  {L.}~\bibnamefont {Bi}}, \ and\ \bibinfo {author} {\bibfnamefont
  {H.}~\bibnamefont {Yu}},\ }\href {\doibase 10.1063/1.5093265} {\bibfield
  {journal} {\bibinfo  {journal} {Applied Physics Letters}\ }\textbf {\bibinfo
  {volume} {114}},\ \bibinfo {pages} {212401} (\bibinfo {year}
  {2019}{\natexlab{a}})}\BibitemShut {NoStop}%
\bibitem [{\citenamefont {Sheng}\ \emph {et~al.}(2020)\citenamefont {Sheng},
  \citenamefont {Liu}, \citenamefont {Chen}, \citenamefont {Wang},
  \citenamefont {Zhang}, \citenamefont {Chen}, \citenamefont {Ma},
  \citenamefont {Liu}, \citenamefont {Tu}, \citenamefont {Nan},\ and\
  \citenamefont {Yu}}]{sheng_spin_2020}%
  \BibitemOpen
  \bibfield  {author} {\bibinfo {author} {\bibfnamefont {L.}~\bibnamefont
  {Sheng}}, \bibinfo {author} {\bibfnamefont {Y.}~\bibnamefont {Liu}}, \bibinfo
  {author} {\bibfnamefont {J.}~\bibnamefont {Chen}}, \bibinfo {author}
  {\bibfnamefont {H.}~\bibnamefont {Wang}}, \bibinfo {author} {\bibfnamefont
  {J.}~\bibnamefont {Zhang}}, \bibinfo {author} {\bibfnamefont
  {M.}~\bibnamefont {Chen}}, \bibinfo {author} {\bibfnamefont {J.}~\bibnamefont
  {Ma}}, \bibinfo {author} {\bibfnamefont {C.}~\bibnamefont {Liu}}, \bibinfo
  {author} {\bibfnamefont {S.}~\bibnamefont {Tu}}, \bibinfo {author}
  {\bibfnamefont {C.-W.}\ \bibnamefont {Nan}}, \ and\ \bibinfo {author}
  {\bibfnamefont {H.}~\bibnamefont {Yu}},\ }\href {\doibase 10.1063/5.0024424}
  {\bibfield  {journal} {\bibinfo  {journal} {Applied Physics Letters}\
  }\textbf {\bibinfo {volume} {117}},\ \bibinfo {pages} {232407} (\bibinfo
  {year} {2020})},\ \bibinfo {note} {publisher: American Institute of
  Physics}\BibitemShut {NoStop}%
\bibitem [{\citenamefont {Sushruth}\ \emph {et~al.}(2020)\citenamefont
  {Sushruth}, \citenamefont {Grassi}, \citenamefont {Ait-Oukaci}, \citenamefont
  {Stoeffler}, \citenamefont {Henry}, \citenamefont {Lacour}, \citenamefont
  {Hehn}, \citenamefont {Bhaskar}, \citenamefont {Bailleul}, \citenamefont
  {Devolder},\ and\ \citenamefont {Adam}}]{sushruth_electrical_2020}%
  \BibitemOpen
  \bibfield  {author} {\bibinfo {author} {\bibfnamefont {M.}~\bibnamefont
  {Sushruth}}, \bibinfo {author} {\bibfnamefont {M.}~\bibnamefont {Grassi}},
  \bibinfo {author} {\bibfnamefont {K.}~\bibnamefont {Ait-Oukaci}}, \bibinfo
  {author} {\bibfnamefont {D.}~\bibnamefont {Stoeffler}}, \bibinfo {author}
  {\bibfnamefont {Y.}~\bibnamefont {Henry}}, \bibinfo {author} {\bibfnamefont
  {D.}~\bibnamefont {Lacour}}, \bibinfo {author} {\bibfnamefont
  {M.}~\bibnamefont {Hehn}}, \bibinfo {author} {\bibfnamefont {U.}~\bibnamefont
  {Bhaskar}}, \bibinfo {author} {\bibfnamefont {M.}~\bibnamefont {Bailleul}},
  \bibinfo {author} {\bibfnamefont {T.}~\bibnamefont {Devolder}}, \ and\
  \bibinfo {author} {\bibfnamefont {J.-P.}\ \bibnamefont {Adam}},\ }\href
  {\doibase 10.1103/PhysRevResearch.2.043203} {\bibfield  {journal} {\bibinfo
  {journal} {Physical Review Research}\ }\textbf {\bibinfo {volume} {2}},\
  \bibinfo {pages} {043203} (\bibinfo {year} {2020})}\BibitemShut {NoStop}%
\bibitem [{\citenamefont {Bailleul}(2013)}]{bailleul_shielding_2013}%
  \BibitemOpen
  \bibfield  {author} {\bibinfo {author} {\bibfnamefont {M.}~\bibnamefont
  {Bailleul}},\ }\href {\doibase 10.1063/1.4829367} {\bibfield  {journal}
  {\bibinfo  {journal} {Applied Physics Letters}\ }\textbf {\bibinfo {volume}
  {103}},\ \bibinfo {pages} {192405} (\bibinfo {year} {2013})}\BibitemShut
  {NoStop}%
\bibitem [{\citenamefont {Talmelli}\ \emph {et~al.}(2020)\citenamefont
  {Talmelli}, \citenamefont {Devolder}, \citenamefont {Träger}, \citenamefont
  {Förster}, \citenamefont {Wintz}, \citenamefont {Weigand}, \citenamefont
  {Stoll}, \citenamefont {Heyns}, \citenamefont {Schütz}, \citenamefont
  {Radu}, \citenamefont {Gräfe}, \citenamefont {Ciubotaru},\ and\
  \citenamefont {Adelmann}}]{talmelli_reconfigurable_2020}%
  \BibitemOpen
  \bibfield  {author} {\bibinfo {author} {\bibfnamefont {G.}~\bibnamefont
  {Talmelli}}, \bibinfo {author} {\bibfnamefont {T.}~\bibnamefont {Devolder}},
  \bibinfo {author} {\bibfnamefont {N.}~\bibnamefont {Träger}}, \bibinfo
  {author} {\bibfnamefont {J.}~\bibnamefont {Förster}}, \bibinfo {author}
  {\bibfnamefont {S.}~\bibnamefont {Wintz}}, \bibinfo {author} {\bibfnamefont
  {M.}~\bibnamefont {Weigand}}, \bibinfo {author} {\bibfnamefont
  {H.}~\bibnamefont {Stoll}}, \bibinfo {author} {\bibfnamefont
  {M.}~\bibnamefont {Heyns}}, \bibinfo {author} {\bibfnamefont
  {G.}~\bibnamefont {Schütz}}, \bibinfo {author} {\bibfnamefont {I.~P.}\
  \bibnamefont {Radu}}, \bibinfo {author} {\bibfnamefont {J.}~\bibnamefont
  {Gräfe}}, \bibinfo {author} {\bibfnamefont {F.}~\bibnamefont {Ciubotaru}}, \
  and\ \bibinfo {author} {\bibfnamefont {C.}~\bibnamefont {Adelmann}},\ }\href
  {\doibase 10.1126/sciadv.abb4042} {\bibfield  {journal} {\bibinfo  {journal}
  {Science Advances}\ }\textbf {\bibinfo {volume} {6}},\ \bibinfo {pages}
  {eabb4042} (\bibinfo {year} {2020})},\ \bibinfo {note} {publisher: American
  Association for the Advancement of Science Section: Research
  Article}\BibitemShut {NoStop}%
\bibitem [{\citenamefont {Demidov}\ and\ \citenamefont
  {Demokritov}(2015)}]{demidov_magnonic_2015}%
  \BibitemOpen
  \bibfield  {author} {\bibinfo {author} {\bibfnamefont {V.~E.}\ \bibnamefont
  {Demidov}}\ and\ \bibinfo {author} {\bibfnamefont {S.~O.}\ \bibnamefont
  {Demokritov}},\ }\href {\doibase 10.1109/TMAG.2014.2388196} {\bibfield
  {journal} {\bibinfo  {journal} {IEEE Transactions on Magnetics}\ }\textbf
  {\bibinfo {volume} {51}},\ \bibinfo {pages} {1} (\bibinfo {year}
  {2015})}\BibitemShut {NoStop}%
\bibitem [{\citenamefont {Bayer}\ \emph {et~al.}(2005)\citenamefont {Bayer},
  \citenamefont {Jorzick}, \citenamefont {Hillebrands}, \citenamefont
  {Demokritov}, \citenamefont {Kouba}, \citenamefont {Bozinoski}, \citenamefont
  {Slavin}, \citenamefont {Guslienko}, \citenamefont {Berkov}, \citenamefont
  {Gorn},\ and\ \citenamefont {Kostylev}}]{bayer_spin-wave_2005}%
  \BibitemOpen
  \bibfield  {author} {\bibinfo {author} {\bibfnamefont {C.}~\bibnamefont
  {Bayer}}, \bibinfo {author} {\bibfnamefont {J.}~\bibnamefont {Jorzick}},
  \bibinfo {author} {\bibfnamefont {B.}~\bibnamefont {Hillebrands}}, \bibinfo
  {author} {\bibfnamefont {S.~O.}\ \bibnamefont {Demokritov}}, \bibinfo
  {author} {\bibfnamefont {R.}~\bibnamefont {Kouba}}, \bibinfo {author}
  {\bibfnamefont {R.}~\bibnamefont {Bozinoski}}, \bibinfo {author}
  {\bibfnamefont {A.~N.}\ \bibnamefont {Slavin}}, \bibinfo {author}
  {\bibfnamefont {K.~Y.}\ \bibnamefont {Guslienko}}, \bibinfo {author}
  {\bibfnamefont {D.~V.}\ \bibnamefont {Berkov}}, \bibinfo {author}
  {\bibfnamefont {N.~L.}\ \bibnamefont {Gorn}}, \ and\ \bibinfo {author}
  {\bibfnamefont {M.~P.}\ \bibnamefont {Kostylev}},\ }\href {\doibase
  10.1103/PhysRevB.72.064427} {\bibfield  {journal} {\bibinfo  {journal}
  {Physical Review B}\ }\textbf {\bibinfo {volume} {72}},\ \bibinfo {pages}
  {064427} (\bibinfo {year} {2005})}\BibitemShut {NoStop}%
\bibitem [{\citenamefont {Vlaminck}\ and\ \citenamefont
  {Bailleul}(2008)}]{vlaminck_current-induced_2008}%
  \BibitemOpen
  \bibfield  {author} {\bibinfo {author} {\bibfnamefont {V.}~\bibnamefont
  {Vlaminck}}\ and\ \bibinfo {author} {\bibfnamefont {M.}~\bibnamefont
  {Bailleul}},\ }\href {\doibase 10.1126/science.1162843} {\bibfield  {journal}
  {\bibinfo  {journal} {Science}\ }\textbf {\bibinfo {volume} {322}},\ \bibinfo
  {pages} {410} (\bibinfo {year} {2008})}\BibitemShut {NoStop}%
\bibitem [{\citenamefont {Vlaminck}\ and\ \citenamefont
  {Bailleul}(2010)}]{vlaminck_spin-wave_2010}%
  \BibitemOpen
  \bibfield  {author} {\bibinfo {author} {\bibfnamefont {V.}~\bibnamefont
  {Vlaminck}}\ and\ \bibinfo {author} {\bibfnamefont {M.}~\bibnamefont
  {Bailleul}},\ }\href {\doibase 10.1103/PhysRevB.81.014425} {\bibfield
  {journal} {\bibinfo  {journal} {Physical Review B}\ }\textbf {\bibinfo
  {volume} {81}},\ \bibinfo {pages} {014425} (\bibinfo {year}
  {2010})}\BibitemShut {NoStop}%
\bibitem [{\citenamefont {Gladii}\ \emph {et~al.}(2016)\citenamefont {Gladii},
  \citenamefont {Collet}, \citenamefont {Garcia-Hernandez}, \citenamefont
  {Cheng}, \citenamefont {Xavier}, \citenamefont {Bortolotti}, \citenamefont
  {Cros}, \citenamefont {Henry}, \citenamefont {Kim}, \citenamefont {Anane},\
  and\ \citenamefont {Bailleul}}]{gladii_spin_2016}%
  \BibitemOpen
  \bibfield  {author} {\bibinfo {author} {\bibfnamefont {O.}~\bibnamefont
  {Gladii}}, \bibinfo {author} {\bibfnamefont {M.}~\bibnamefont {Collet}},
  \bibinfo {author} {\bibfnamefont {K.}~\bibnamefont {Garcia-Hernandez}},
  \bibinfo {author} {\bibfnamefont {C.}~\bibnamefont {Cheng}}, \bibinfo
  {author} {\bibfnamefont {S.}~\bibnamefont {Xavier}}, \bibinfo {author}
  {\bibfnamefont {P.}~\bibnamefont {Bortolotti}}, \bibinfo {author}
  {\bibfnamefont {V.}~\bibnamefont {Cros}}, \bibinfo {author} {\bibfnamefont
  {Y.}~\bibnamefont {Henry}}, \bibinfo {author} {\bibfnamefont {J.-V.}\
  \bibnamefont {Kim}}, \bibinfo {author} {\bibfnamefont {A.}~\bibnamefont
  {Anane}}, \ and\ \bibinfo {author} {\bibfnamefont {M.}~\bibnamefont
  {Bailleul}},\ }\href {\doibase 10.1063/1.4952447} {\bibfield  {journal}
  {\bibinfo  {journal} {Applied Physics Letters}\ }\textbf {\bibinfo {volume}
  {108}},\ \bibinfo {pages} {202407} (\bibinfo {year} {2016})}\BibitemShut
  {NoStop}%
\bibitem [{\citenamefont {Bhaskar}\ \emph {et~al.}(2020)\citenamefont
  {Bhaskar}, \citenamefont {Talmelli}, \citenamefont {Ciubotaru}, \citenamefont
  {Adelmann},\ and\ \citenamefont {Devolder}}]{bhaskar_backward_2020}%
  \BibitemOpen
  \bibfield  {author} {\bibinfo {author} {\bibfnamefont {U.~K.}\ \bibnamefont
  {Bhaskar}}, \bibinfo {author} {\bibfnamefont {G.}~\bibnamefont {Talmelli}},
  \bibinfo {author} {\bibfnamefont {F.}~\bibnamefont {Ciubotaru}}, \bibinfo
  {author} {\bibfnamefont {C.}~\bibnamefont {Adelmann}}, \ and\ \bibinfo
  {author} {\bibfnamefont {T.}~\bibnamefont {Devolder}},\ }\href {\doibase
  10.1063/1.5125751} {\bibfield  {journal} {\bibinfo  {journal} {Journal of
  Applied Physics}\ }\textbf {\bibinfo {volume} {127}},\ \bibinfo {pages}
  {033902} (\bibinfo {year} {2020})}\BibitemShut {NoStop}%
\bibitem [{\citenamefont {Demidov}\ \emph {et~al.}(2008)\citenamefont
  {Demidov}, \citenamefont {Demokritov}, \citenamefont {Rott}, \citenamefont
  {Krzysteczko},\ and\ \citenamefont {Reiss}}]{demidov_mode_2008}%
  \BibitemOpen
  \bibfield  {author} {\bibinfo {author} {\bibfnamefont {V.~E.}\ \bibnamefont
  {Demidov}}, \bibinfo {author} {\bibfnamefont {S.~O.}\ \bibnamefont
  {Demokritov}}, \bibinfo {author} {\bibfnamefont {K.}~\bibnamefont {Rott}},
  \bibinfo {author} {\bibfnamefont {P.}~\bibnamefont {Krzysteczko}}, \ and\
  \bibinfo {author} {\bibfnamefont {G.}~\bibnamefont {Reiss}},\ }\href
  {\doibase 10.1103/PhysRevB.77.064406} {\bibfield  {journal} {\bibinfo
  {journal} {Physical Review B}\ }\textbf {\bibinfo {volume} {77}},\ \bibinfo
  {pages} {064406} (\bibinfo {year} {2008})}\BibitemShut {NoStop}%
\bibitem [{\citenamefont {Brächer}\ \emph {et~al.}(2017)\citenamefont
  {Brächer}, \citenamefont {Boulle}, \citenamefont {Gaudin},\ and\
  \citenamefont {Pirro}}]{bracher_creation_2017}%
  \BibitemOpen
  \bibfield  {author} {\bibinfo {author} {\bibfnamefont {T.}~\bibnamefont
  {Brächer}}, \bibinfo {author} {\bibfnamefont {O.}~\bibnamefont {Boulle}},
  \bibinfo {author} {\bibfnamefont {G.}~\bibnamefont {Gaudin}}, \ and\ \bibinfo
  {author} {\bibfnamefont {P.}~\bibnamefont {Pirro}},\ }\href {\doibase
  10.1103/PhysRevB.95.064429} {\bibfield  {journal} {\bibinfo  {journal}
  {Physical Review B}\ }\textbf {\bibinfo {volume} {95}},\ \bibinfo {pages}
  {064429} (\bibinfo {year} {2017})},\ \bibinfo {note} {publisher: American
  Physical Society}\BibitemShut {NoStop}%
\bibitem [{\citenamefont {Slavin}, \citenamefont {Demokritov},\ and\
  \citenamefont {Hillebrands}(2002)}]{slavin_nonlinear_2002}%
  \BibitemOpen
  \bibfield  {author} {\bibinfo {author} {\bibfnamefont {A.~N.}\ \bibnamefont
  {Slavin}}, \bibinfo {author} {\bibfnamefont {S.~O.}\ \bibnamefont
  {Demokritov}}, \ and\ \bibinfo {author} {\bibfnamefont {B.}~\bibnamefont
  {Hillebrands}},\ }in\ \href {\doibase 10.1007/3-540-40907-6_2} {\emph
  {\bibinfo {booktitle} {Spin {Dynamics} in {Confined} {Magnetic} {Structures}
  {I}}}},\ \bibinfo {series and number} {Topics in {Applied} {Physics}},\
  \bibinfo {editor} {edited by\ \bibinfo {editor} {\bibfnamefont
  {B.}~\bibnamefont {Hillebrands}}\ and\ \bibinfo {editor} {\bibfnamefont
  {K.}~\bibnamefont {Ounadjela}}}\ (\bibinfo  {publisher} {Springer},\ \bibinfo
  {address} {Berlin, Heidelberg},\ \bibinfo {year} {2002})\ pp.\ \bibinfo
  {pages} {35--64}\BibitemShut {NoStop}%
\bibitem [{\citenamefont {Silva}\ \emph {et~al.}(1999)\citenamefont {Silva},
  \citenamefont {Lee}, \citenamefont {Crawford},\ and\ \citenamefont
  {Rogers}}]{silva_inductive_1999}%
  \BibitemOpen
  \bibfield  {author} {\bibinfo {author} {\bibfnamefont {T.~J.}\ \bibnamefont
  {Silva}}, \bibinfo {author} {\bibfnamefont {C.~S.}\ \bibnamefont {Lee}},
  \bibinfo {author} {\bibfnamefont {T.~M.}\ \bibnamefont {Crawford}}, \ and\
  \bibinfo {author} {\bibfnamefont {C.~T.}\ \bibnamefont {Rogers}},\ }\href
  {\doibase 10.1063/1.370596} {\bibfield  {journal} {\bibinfo  {journal}
  {Journal of Applied Physics}\ }\textbf {\bibinfo {volume} {85}},\ \bibinfo
  {pages} {7849} (\bibinfo {year} {1999})},\ \bibinfo {note} {publisher:
  American Institute of Physics}\BibitemShut {NoStop}%
\bibitem [{\citenamefont {Birt}\ \emph {et~al.}(2012)\citenamefont {Birt},
  \citenamefont {An}, \citenamefont {Tsoi}, \citenamefont {Tamaru},
  \citenamefont {Ricketts}, \citenamefont {Wong}, \citenamefont
  {Khalili~Amiri}, \citenamefont {Wang},\ and\ \citenamefont
  {Li}}]{birt_deviation_2012}%
  \BibitemOpen
  \bibfield  {author} {\bibinfo {author} {\bibfnamefont {D.~R.}\ \bibnamefont
  {Birt}}, \bibinfo {author} {\bibfnamefont {K.}~\bibnamefont {An}}, \bibinfo
  {author} {\bibfnamefont {M.}~\bibnamefont {Tsoi}}, \bibinfo {author}
  {\bibfnamefont {S.}~\bibnamefont {Tamaru}}, \bibinfo {author} {\bibfnamefont
  {D.}~\bibnamefont {Ricketts}}, \bibinfo {author} {\bibfnamefont {K.~L.}\
  \bibnamefont {Wong}}, \bibinfo {author} {\bibfnamefont {P.}~\bibnamefont
  {Khalili~Amiri}}, \bibinfo {author} {\bibfnamefont {K.~L.}\ \bibnamefont
  {Wang}}, \ and\ \bibinfo {author} {\bibfnamefont {X.}~\bibnamefont {Li}},\
  }\href {\doibase 10.1063/1.4772798} {\bibfield  {journal} {\bibinfo
  {journal} {Applied Physics Letters}\ }\textbf {\bibinfo {volume} {101}},\
  \bibinfo {pages} {252409} (\bibinfo {year} {2012})}\BibitemShut {NoStop}%
\bibitem [{\citenamefont {Chen}\ \emph
  {et~al.}(2019{\natexlab{b}})\citenamefont {Chen}, \citenamefont {Yu},
  \citenamefont {Liu}, \citenamefont {Liu}, \citenamefont {Madami},
  \citenamefont {Shen}, \citenamefont {Zhang}, \citenamefont {Tu},
  \citenamefont {Alam}, \citenamefont {Xia}, \citenamefont {Wu}, \citenamefont
  {Gubbiotti}, \citenamefont {Blanter}, \citenamefont {Bauer},\ and\
  \citenamefont {Yu}}]{chen_excitation_2019}%
  \BibitemOpen
  \bibfield  {author} {\bibinfo {author} {\bibfnamefont {J.}~\bibnamefont
  {Chen}}, \bibinfo {author} {\bibfnamefont {T.}~\bibnamefont {Yu}}, \bibinfo
  {author} {\bibfnamefont {C.}~\bibnamefont {Liu}}, \bibinfo {author}
  {\bibfnamefont {T.}~\bibnamefont {Liu}}, \bibinfo {author} {\bibfnamefont
  {M.}~\bibnamefont {Madami}}, \bibinfo {author} {\bibfnamefont
  {K.}~\bibnamefont {Shen}}, \bibinfo {author} {\bibfnamefont {J.}~\bibnamefont
  {Zhang}}, \bibinfo {author} {\bibfnamefont {S.}~\bibnamefont {Tu}}, \bibinfo
  {author} {\bibfnamefont {M.~S.}\ \bibnamefont {Alam}}, \bibinfo {author}
  {\bibfnamefont {K.}~\bibnamefont {Xia}}, \bibinfo {author} {\bibfnamefont
  {M.}~\bibnamefont {Wu}}, \bibinfo {author} {\bibfnamefont {G.}~\bibnamefont
  {Gubbiotti}}, \bibinfo {author} {\bibfnamefont {Y.~M.}\ \bibnamefont
  {Blanter}}, \bibinfo {author} {\bibfnamefont {G.~E.~W.}\ \bibnamefont
  {Bauer}}, \ and\ \bibinfo {author} {\bibfnamefont {H.}~\bibnamefont {Yu}},\
  }\href {\doibase 10.1103/PhysRevB.100.104427} {\bibfield  {journal} {\bibinfo
   {journal} {Physical Review B}\ }\textbf {\bibinfo {volume} {100}},\ \bibinfo
  {pages} {104427} (\bibinfo {year} {2019}{\natexlab{b}})}\BibitemShut
  {NoStop}%
\end{thebibliography}
%

\end{document}